\documentclass[11pt,letterpaper]{article}
\pdfoutput=1

\usepackage[margin=1in]{geometry}

\usepackage[small,compact]{titlesec}

\RequirePackage[normalem]{ulem} 
\RequirePackage{color}\definecolor{RED}{rgb}{1,0,0}\definecolor{BLUE}{rgb}{0,0,1} 

\usepackage{times}
\usepackage{amsmath}
\usepackage{graphicx}
\usepackage{subfigure}
\usepackage{url}
\usepackage{mdwlist}
\usepackage{cite}
\usepackage{color}
\usepackage{soul}          

\newcommand{\tkey}{\mathit{key}}
\newcommand{\xor}{\mathit{XOR}}
\newcommand{\visited}{\mathit{visited}}
\newcommand{\increment}{\mathit{increment}}
\newcommand{\amark}{\mathit{mark}}

\usepackage{algorithmic}
\usepackage[algonl, ruled]{algorithm2e}

\usepackage{balance}

\usepackage{hyperref}
\hypersetup{colorlinks=false, pdfborder= 0 0 0}

\usepackage{xspace}

\usepackage{tkz-graph}

\newcommand{\scalefac}{0.45}
\newcommand{\sys}{ReDS\xspace}
\newcommand{\hsys}{Halo-ReDS\xspace}
\newcommand{\ksys}{Kad-ReDS\xspace}

\newcommand{\comment}[1]{}
\newcommand{\sharedrep}{Drop-off\xspace}

\newcommand{\putk}{\mathsf{put}}
\newcommand{\getk}{\mathsf{get}}
\newcommand{\key}{\mathsf{key}}
\newcommand{\object}{\mathsf{object}}
\newcommand{\lookup}{\mathsf{lookup}}

\newcommand{\paragraphX}[1]{\vskip 4pt \noindent \textbf{#1} \hskip .05in}

\newenvironment{smitemize}
  {\begin{list}{$\bullet$}
     {\setlength{\parsep}{1pt}
      \setlength{\leftmargin}{8pt}
      \setlength{\topsep}{2pt}
      \setlength{\itemsep}{0pt}}}
  {\end{list}}

\begin{document}

\title{ReDS: A Framework for Reputation-Enhanced DHTs}

\author{Ruj~Akavipat$^{\dag}$, Mahdi~N.~Al-Ameen$^{\S}$, Apu~Kapadia$^{\ddag}$, Zahid~Rahman$^{\ddag}$, \\Roman~Schlegel$^{\ddag,\sharp}$, Matthew~Wright$^{\S}$\\\\
\begin{tabular}{ccc}
$^{\dag}$Dept. of Computer Engineering & ~ & $^{\S}$Dept. of Computer Science and Engineering\\
Mahidol University & & University of Texas at Arlington\\
Salaya, Nakornpathom, Thailand & & Arlington, TX, USA  \\
akavipat@gmail.com & & mahdi.al-ameen@mavs.uta.edu\\
 & & mwright@cse.uta.edu\\\\
 $^{\ddag}$School of Informatics and Computing & ~ &  $^{\sharp}$Department of Computer Science\\
 Indiana University  & ~ & City University of Hong Kong\\
  Bloomington, IN, USA  & ~ & Kowloon Tong, Kowloon, Hong Kong\\
  rahmanmd@indiana.edu & ~ & sschlegel2@student.cityu.edu.hk\\
  kapadia@indiana.edu\\\\
\end{tabular}\\ \\
  }



\maketitle

\sloppy





\thispagestyle{empty}
\begin{abstract}
Distributed Hash Tables (DHTs) such as Chord and Kademlia offer
an efficient solution for locating resources in peer-to-peer networks.
Unfortunately, malicious nodes along a lookup path can easily subvert
such queries.
Several systems, including Halo (based on Chord) and Kad (based on Kademlia),
mitigate such attacks by using a combination of redundancy and diversity in the paths taken by redundant lookup queries. 
Much greater assurance can be provided, however. We describe Reputation for Directory Services (ReDS), a framework
for enhancing  lookups in redundant DHTs by tracking how well other nodes
service lookup requests. We describe how the ReDS technique can be
applied to virtually any redundant DHT including Halo and Kad. 
We also study the collaborative identification and removal of
bad lookup paths in a way that  does not rely on the sharing
of reputation scores~--- we show that such sharing is vulnerable to
attacks that make it unsuitable for most applications of ReDS. Through
extensive simulations  we demonstrate that ReDS
improves lookup success rates for Halo and Kad by 80\% or more over a wide range of conditions, even against strategic attackers
attempting to game their reputation scores and in the presence of node
churn. 
\end{abstract}
\newpage

\section{Introduction}
\label{sec:intro}

Over the past several years peer-to-peer (P2P) systems have been gaining popularity and mainstream
acceptance. For example, Skype, the popular P2P-based system,
had 42 million concurrent online users in September
2012.\footnote{\url{http://skypenumerology.blogspot.com/}}
BitTorrent,\footnote{\url{http://www.bittorrent.com/}}
Akamai,\footnote{Akamai NetSession Interface
  \url{http://www.akamai.com/html/misc/akamai_client/netsession_interface_overview.html}}
and even botnets are large P2P systems that must achieve decentralized
coordination to locate resources. For example, in Skype one must be able
to locate the current IP addresses of contacts, and in a distributed
storage system one must be able to locate the IP address of a node
hosting a particular file. One class of solutions called {\em
  distributed hash tables} ({\em DHTs}) maps resources onto nodes in the P2P
network and provides a put-get abstraction where resources can be stored
(put) in the network and subsequently retrieved (get). The key idea in
DHTs is that each peer maintains a routing table with only a few entries
and yet any resource can be located by routing queries through a few
nodes, where ``few'' usually corresponds to a number logarithmic in the
number of nodes in the network. Chord~\cite{chord}, CAN~\cite{CAN},
Pastry~\cite{pastry}, and Kademlia~\cite{kad} are examples of DHTs with
these properties.


DHTs provide several important properties, such as scalable location of
nodes and services, but do not protect against malicious peers
manipulating the resource locate operations ({\em lookups}). For example, an attacker
may want to undermine the system's operations by providing fake
lookup results for non-existent peers or to make his own peers the end
point of lookups so as to pollute the network's files and services. Such
an attacker can easily manipulate much of the system's activity. In
Chord, for example, about 10\% of malicious nodes in the network can
subvert more than 50\% of the searches~\cite{halo}. Halo is a system that exploits the
deterministic mapping of routing-table entries to nodes in Chord to provide a `high-assurance locate' in
Chord through redundant searches~\cite{halo}. Several other DHTs such as
Salsa~\cite{salsa}, Cyclone~\cite{cyclone}, and NISAN~\cite{nisan} also
utilize redundant searching to tolerate malicious nodes in the network. 
Kad (based on  Kademlia) is an example of a non-deterministic DHT that also incorporates redundancy into the protocol; routing-table entries are a function of nodes encountered in the system and are not easily predictable.


While all these techniques are able to improve the success of lookups by
a combination of redundancy and diversity of the redundant lookup paths,
they still allow a non-trivial failure rate while incurring substantial
overhead for redundancy.
For example, Halo still has a failure rate as high as 5--6\% for 20\%
malicious nodes utilizing a logarithmic number of redundant lookups in
the size of the network (e.g., 13 lookups in a network of 10,000 nodes).
We show that  
Kad  has a non-trivial failure rate of 17--21\% 
with only 10\% of malicious nodes even with a high level of redundancy. 
In this paper we investigate an approach to improve DHT lookups in
malicious environments. The central observation of our technique is a
simple one: if a node uses redundant lookups and tracks which nodes gave
accurate results, then it can use this information to improve the success
rate of lookups that traverse it.


Our design approach,  Reputation for Directory Services (ReDS),
includes two novel features. First, the querying node uses the
redundancy in the lookups and structure of the DHT to infer the honesty
and reliability of nodes throughout the lookup path, even though direct
observation is not possible.
%
Second, the peers employ `collaborative boosting' in which each
node involved in the lookup routing can improve the success of the route
by picking the next hop based on a form of constrained local boosting
(so as not to inflate the path length significantly).

We note that quite a bit of work has been done on P2P reputation
systems (Hoffman et al. provide an extensive survey~\cite{hoffman09survey}). However, this work mainly addresses the
`free rider problem' in which some users unfairly use resources
provided by peers without providing any resources themselves. This issue
is  orthogonal to our work, which instead leverages reputation
to detect malicious behavior that aims to undermine DHT routing. While
we are able to leverage some of the findings of other work on reputation
systems, identifying malicious behavior in the DHT routing layer
presents unique design challenges that we address.

\paragraphX{Contributions.} Preliminary results on ReDS were published as a work-in-progress paper examining ReDS in the context of Salsa~\cite{reds-csiirw10}  and a workshop paper examining Halo-ReDS under a limited adversary model~\cite{reds-wits10}.
Here we make the following additional significant contributions:


\begin{smitemize}
\item We show  that reputation can be applied to a variety of
  redundant DHT-based directory services to improve lookup success
  rates. We specifically describe {\em Halo-ReDS} and {\em Kad-ReDS},
  which are implementations for Halo (a deterministic DHT) and Kad (a
  non-deterministic DHT).


 
\item Building on our approach of using a \emph{reputation tree} to make statistical inferences about where malicious nodes reside in the DHT, we show how nodes along a lookup path can make use of their local reputation trees for
  \emph{collaborative boosting}.  We show
  a dramatic improvement in success rates for this mode.


\item Through analysis and simulation we study the behavior of Halo-ReDS
  and Kad-ReDS under adaptive adversaries who attack only some fraction
  of the time in an effort to game the reputation system.
  In particular we show that attackers are limited to attacking at a  
  low, ineffective rate.
%
  

\item We evaluate the performance of  Halo-ReDS and Kad-ReDS under
   churn and show that while adaptive inference  suffers
  with churn, collaborative ReDS is more robust.

\item We examine the possibility of sharing reputation scores and 
show how such sharing can be attacked
  though slandering and self-promotion
  attacks. Further, we identify a new `use-based' attack on shared reputation that
  would greatly undermine most ReDS systems, leading us to recommend
  using only first-hand observations.
\end{smitemize}



\section{System and Attack Models}
\label{sec:model}
%
%
\subsection{System model}

DHTs support a distributed implementation of $\putk$ and $\getk$
operations, where objects are placed on nodes in the peer-to-peer
network and are indexed by $\key$s. By using operations such as
$\putk(\key, \object)$ and $\getk(\key)$, objects can be inserted into
and retrieved from the DHT. In both operations, the DHT must first map
the key to a particular \emph{owner} node $o$ in the system. Once owner
$o$ has been located, the resource is inserted or retrieved from $o$. We describe the rest of the system model in the context of Halo and Kad.

\paragraphX{Halo and Chord.}
In Chord, nodes are assigned to a virtual address space organized in a
ring. For example, the address space could correspond to the output of
SHA-1, and the next address after $2^{160}-1$ is 0 again. IDs can be
issued to nodes via a central authority along with a
certificate~\cite{castro02secp2p}.
Resources such as files can be assigned virtual addresses (the
resource's $\key$) based on the hash values of their filenames. A
resource's owner is the {\em clockwise-closest} in the virtual address
space, i.e. the node with the lowest ID greater than the target ID,
modulo the size of the ID space. 
Each node maintains a routing table of nodes called
``fingers'', which are at exponentially increasing distances from itself. 
When a node receives a locate request for a target key $t$, it redirects
the query to the closest finger to $t$. This process results in efficient lookups 
with $O(\log n)$ hops requiring only $O(\log n)$ storage at each node, where
$n$ is the total number of nodes in the DHT.

While Chord has good stability properties under independent node
failures, lookups are easily subverted. Simply adding redundancy to
lookups does not help very much, as lookups often converge to the same
nodes.
Halo makes the
observation that each node $v$ occurs in $O(\log n)$ other nodes' finger
tables~\cite{halo}. These nodes are called the ``knuckles'' of $v$.
Searching for those knuckles
instead of the
actual target effectively disentangles the redundant searches.  Note that because of the
clockwise-closest relation, if a redundant search yields multiple
candidates for the target's owner, then the closest one (that is alive) is
picked. Thus, as long as one of the lookups in a redundant search
returns the correct answer, then the correct owner is obtained.





\paragraphX{Kad.}
\label{kad}
Kad is a widely deployed DHT based on Kademlia~\cite{kad}. 
Distances in the Kad ID space are measured using the XOR of two
IDs and taking the output as an integer. 
In Kad, each node maintains a routing table comprising of `$k$-buckets' for each exponentially increasing interval of ID space from the node.  Each $k$-bucket includes up to $k$
nodes from the corresponding ID range and are dynamically populated by new nodes encountered in each $\putk$-$\getk$ operation, resulting in the non-determinism of routing table entries as compared to other deterministic DHTs such as Chord. More specifically, the $j$-th $k$-bucket of a node contains learned nodes for which it shares the first $j$ bits of the
ID and has a different $j+1$-st bit. If a $k$-bucket is full, then the least recently seen node is evicted to make space for a new node.


Kad lookups proceed {\em iteratively}, where each node contacts $\alpha$ nodes at each step and receives the $\beta$ closest results from each of them. A short list of $k$ nodes is maintained by the querying node and the list is updated with the $\alpha \times \beta$ results returned at every step.
At the next step the
querying node contacts the closest $\alpha$ unqueried nodes drawn from the short list. Kad
ensures $O(\log n)$ lookup steps by moving at least one bit closer to the target ID with each iteration.

In Kad, a `resource' is stored on $r$ different nodes (called {\em replica roots}) around the key such that their Kad ID falls within a certain distance called {\em search tolerance}, $\delta$. Typically $r=10$ and $\delta$ is such that Kad ID of a replica root agrees at least in the first $8$ bits with the key.

\subsection{Attack model}\label{sec:model:attack}

Malicious nodes in the system may attempt to subvert lookup operations,
e.g., by dropping or misdirecting lookups. 
The adversary's goal could be to cause peers to use attacker-controlled
nodes, e.g., as a way to spread
disinformation, spam, or malware.
 Adversaries may also seek to censor access to content through
denial of service or degradation of service attacks in which lookup
results lead to invalid or incorrect nodes. The attacker's most
effective strategy to achieve these ends in a P2P network
is to control a large number of peers in the system (or virtual
peers by controlling its location in the address space). To prevent the
number of malicious peers from growing without bound,
social-network-based anti-Sybil techniques such as
SybilInfer~\cite{sybilinfer} and SybilLimit~\cite{sybillimit} may be
used. However, we expect that the attacker may be able to inject a
constant fraction of the total number of peers into the network without
detection through social engineering. 

Such an attacker would both try to  manipulate lookup results as
well as try to deceive any attempt at using reputation or malicious node
detection. Thus, our system design has to take both types of attacks into
account.
We assume  lookup operations going through a malicious node will be
manipulated by the malicious node to map the key to the closest
malicious node instead of the actual owner. Furthermore, we also assume
 the attackers can coordinate and choose to attack only a fraction
of the time to evade detection. Thus, for an attack rate $a$ (measured as
a probability), adversaries will compromise a particular $\lookup(t)$
for target $t$ with probability $a$. We assume powerful adversaries who
can exchange information in real time and flag $t$ as a target that
should be attacked or not.
%

A standard assumption we make in our analysis is that malicious nodes
cannot control their placement in the ring and thus malicious nodes are
distributed uniformly at random in the address space. The use of a SHA-1
hash can achieve such a distribution as long as the attackers cannot
control the value of their hashed identity and cannot own a large number
of identities. Thus, we must assume that all peers have some identifier
(such as a PKI credential) issued by a trusted provider. Furthermore,
peers can verify the virtual addresses of nodes by checking
signatures, e.g., Myrmic~\cite{myrmic-tr} provides
such assurances for Kademlia.


For Halo we assume that  `control' lookups for maintaining routing tables use  high enough redundancy to ensure minimal chances of
attackers gaining any extra influence in the system. Kad's routing tables are updated during regular lookup operations, and we show how our approach effectively limits routing table pollution.





\section{\sys Design}
\label{sec:reds}

We propose to augment DHTs like Halo and Kad so that nodes can utilize
the successes and failures of individual lookups in a redundant search
to infer and avoid malicious nodes in the DHT. \sys can then better direct searches in two steps: 1) the
originator of a lookup picks the best possible start node(s) (`local boosting') and 2) each node involved in
the lookup process can avoid malicious fingers by selecting alternative
fingers  (`collaborative boosting'). 

\subsection{Halo-ReDS}
\label{sec:halo-reds}


In previous work, we describe a local boosting algorithm called
\emph{A-Boost} (for \emph{adaptive boost})~\cite{reds-wits10}, where
local observations are used to predict lookup performance as far along
an entire lookup path as possible. The general idea is to attempt to
infer the locations of malicious nodes in the path, and A-Boost makes
such determinations based
on the amount of reputation information collected for that path. The
entire set of reputation information is called the {\em reputation
  tree}. We refer the reader to the paper for more information on
A-Boost~\cite{reds-wits10}.

We now describe a simple technique by which nodes can improve the
success of a lookup by collaboratively using their locally collected
reputation scores to improve finger selection for the next hop. In Kad
we already expect to have multiple fingers in each $k$-bucket. For
Halo-ReDS we augment Halo nodes by adding the concept of $k$-buckets,
one for each location where a finger would be in the Chord ring. In
particular the $k$-bucket for a given finger's key ($v+2^{i}$) includes
that key's owner (the original Chord finger) and the $k-1$ predecessors
of that node.
For collaborative boosting each node maintains a reputation tree, as in
A-Boost. When the closest finger for a target key $t$ is requested by
node $u$, $u$ checks the A-Boost scores of the nodes in the $k$-bucket
closest to $t$ and selects the best finger in the bucket for that
request. Thus, the failure rate at each hop in the lookup is expected to
drop significantly because all nodes in the $k$-bucket must be malicious
to subvert a lookup.

In a Chord lookup the lookup locates the predecessor first and asks it
to return the successor (the target node). Thus, a malicious predecessor
can still subvert a lookup for its successor because it controls
\emph{all} lookups for that successor. To alleviate this problem, we
assume that each node knows $k'$ additional successors ($k'+1$ in total)
so that the last hop is short-circuited as long as the lookup reaches
the $k'$-vicinity of the target.


Another issue in Halo is that selecting the predecessors of a finger for
a $k$-bucket means that the lookup may not get as close to the target at
each hop, thereby increasing the lookup cost. Since each hop may regress
by at most $k$ nodes at each hop, the average number of nodes between
the current hop and the target node after hop $i$ is at most $n/2^{i} -
k/2^{i-1} + 2k$.
Therefore, assuming $k < \log n$, we have that after $O(\log n)$ hops,
there are at most $2k+1$ nodes between the target and the current
hop. As long as $k' \geq 2k+1$, short-circuiting means that this will
take one additional hop compared to regular Chord. In our system, we use
$k=2$ and $k'=8$, so for 1,000 nodes this constraint is satisfied (and
thus for larger networks too). We experimentally validated this analysis
and found that for a 1,000-node network, path lengths for Halo-\sys
increased by only 0.15 hops on average compared to regular Halo.

%

\subsection{Kad-ReDS.}  
\label{sec:kad-reds}

Because of the non-deterministic nature of Kad and the dynamism of
$k$-bucket entries, we follow a different approach for maintaining
reputation in Kad-\sys as described next.

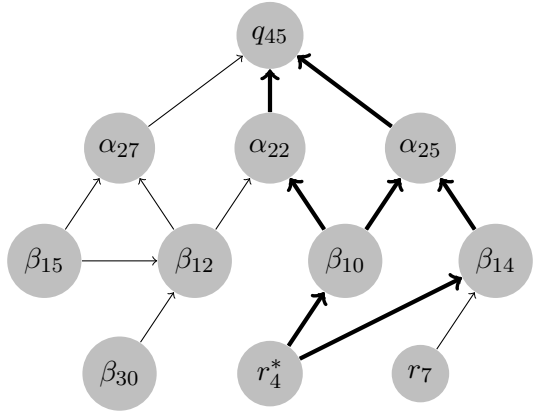
\begin{figure}
\centering
\scalebox{1}
{
\begin{tikzpicture}
\tikzstyle{vertex}=[circle,fill=black!25];
\tikzset{EdgeStyle/.style = {->}}

\node[vertex] (Q) at (0,0) {$q_{45}$};

\node[vertex] (A11) at (-3/1.5,-1.5/1) {$\alpha_{27}$};
\node[vertex] (A12) at (0,-1.5/1) {$\alpha_{22}$};
\node[vertex] (A13) at (3/1.5,-1.5/1) {$\alpha_{25}$};

\node[vertex] (B21) at (-4.5/1.5,-3/1) {$\beta_{15}$};
\node[vertex] (B22) at (-1.5/1.5,-3/1) {$\beta_{12}$};
\node[vertex] (B23) at (1.5/1.5,-3/1) {$\beta_{10}$};
\node[vertex] (B24) at (4.5/1.5,-3/1) {$\beta_{14}$};

\node[vertex] (B31) at (-3/1.5,-4.5/1) {$\beta_{30}$};
\node[vertex] (B32) at (0,-4.5/1) {$r_{4}^{*}$};
\node[vertex] (B33) at (3/1.5,-4.5/1) {$r_{7}$};

\draw[->] (A11) edge (Q);	
\draw[->] (A12) edge (Q);			
\draw[->] (A13) edge (Q);

\draw[->] (B22) edge (A11);	
\draw[->] (B21) edge (B22);	
\draw[->] (B22) edge (A12);
\draw[->] (B23) edge (A12);
\draw[->] (B23) edge (A13);
\draw[->] (B24) edge (A13);

\draw[->] (B21) edge (A11);		
\draw[->] (B31) edge (B22);	
\draw[->] (B32) edge (B23);	
\draw[->] (B32) edge (B24);	
\draw[->] (B33) edge (B24);	

\tikzset{EdgeStyle/.append style= {ultra thick,
 double = orange,
double distance = 1pt}}
\draw[EdgeStyle] (B32) edge (B23);
\draw[EdgeStyle] (B32) edge (B24);	
\draw[EdgeStyle] (B24) edge (A13);
\draw[EdgeStyle] (B23) edge (A13);
\draw[EdgeStyle] (B23) edge (A12);
\draw[EdgeStyle] (A13) edge (Q);
\draw[EdgeStyle] (A12) edge (Q);

\end{tikzpicture}
}
\caption{Lookup graph for a lookup initiated by node $q_{45}$ for $\tkey$. Subscripts denote the $\xor$ distance of the node from the
  lookup target $\tkey$, i.e. $\xor(q_{45}, \tkey) = 45$. Edges from a
  node are directed toward the node from which they are returned during
  the lookup process. Three successful paths from a correct replica root
  $r_{4}$ are shown in bold.} \label{fig:graph}
\end{figure}

\emph{Building the lookup graph.} In Kad-\sys the querying node keeps
track of the lookup paths that are used to locate the target. As a
lookup operation proceeds, the querying node builds a lookup graph with
itself as a vertex as shown in Figure~\ref{fig:graph}. For each of the $\beta$ results, a directed edge is constructed to the intermediate queried node returning that result.
%
%
We note that duplicate nodes
can be returned by different queried nodes during a lookup step, e.g.,
 $\beta_{10}$ in Figure~\ref{fig:graph} is returned by two
different queried nodes $\alpha_{22}$ and $\alpha_{25}$.
 Cycles are also possible, as a node may  return an ancestor.
The querying node incrementally builds such a graph by using
Algorithm~\ref{algo:ldag} at each step of the lookup. For
example, in the next iteration the querying node $q_{45}$ selects the
closest $\alpha$ nodes $\beta_{10}$, $\beta_{12}$, and $\beta_{14}$ to
expand next. We also note that the algorithm is initialized by constructing 
$\alpha$ edges toward the querying node from each of the $\alpha$ nodes it
selects to query at the first step of the lookup.

\begin{algorithm}[t]
  \small
 \SetKwInOut{Input}{input}\SetKwInOut{Output}{output}
 
 \Input{$q$ : Querying node\\
        $i$ : Current lookup step\\
        ${\alpha_{i}}$ : a queried node at step $i$\\
        $N_{\alpha}(1:\beta)$ : $\beta$ result nodes returned by ${\alpha_{i}}$\\
        $G_{i-1}\langle V,E\rangle$ : Lookup graph at step $(i-1)$}
 \Output{$G_{i}\langle V,E\rangle$ at step $i$}
 \BlankLine
 \If{i = 0}{ 
 	 Add vertices $q$ and ${\alpha_{i}}$ in $V(G_{i-1})$\;
 	 Add a directed edge $e=({\alpha_{i}, q})$, from ${\alpha_{i}}$ 
           to $q$ in $E(G_{i-1})$\;
	 }
 \BlankLine
 \For{each node ${\beta}_{j}$ in $N_{\alpha}$}{
 	\If{${\beta}_{j}$ is not a vertex in $G_{i-1}$}
           {Add vertex ${\beta}_{j}$ in $V(G_{i-1})$\;} 
	 
	 Add a directed edge $e=({\beta_{i}}, {\alpha_{i}})$ from
         $\beta_{i}$ to ${\alpha_{i}}$ in $E(G_{i-1})$\;
 }
 \caption{Building the lookup graph for Kad-\sys}
 \label{algo:ldag}
\end{algorithm}

\emph{Applying reputation scores.} After the search terminates with one
or more replica roots being identified, we mark each lookup path as
being successful if it terminated in finding the {\em closest} replica
root to the key. 
This is done by traversing the lookup graph in depth-first search order
as described in Algorithm~\ref{algo:update-rep-kad}, with the closest
replica root as the starting node $u$.
Algorithm~\ref{algo:update-rep-kad} avoids cycles by keeping track of
nodes visited during the traversal. Each finger appearing on a
successful path is then credited +1 to its reputation score since it was
involved in locating the closest correct replica root. Other nodes in
the graph are not credited. For example in Figure~\ref{fig:graph},
reputation score of the contacts $\beta_{10}$ and $\alpha_{25}$ in the
$k$-bucket of $q_{45}$ are increased by 2 as two successful paths go
through these nodes. Our rationale for locating the closest replica root
is to make Kad-\sys robust against attackers who may attempt to insert a
replica root close to the target key. By crediting nodes for locating
the closest replica root, the attacker is forced to insert itself at a
location closer than any other replica root, which is significantly
harder.

\begin{algorithm}[t]
  \small
\SetKwInOut{Input}{input}\SetKwInOut{Output}{output}

\SetKwFunction{KwFn}{UpdateReputation}
\KwFn{}
\Input{
  $q$ : Querying node\\    
  $G\langle V,E\rangle$ : Final lookup graph\\
  $u$ : A vertex in the graph $G\langle V,E\rangle$\\
  $\visited$ : List of visited nodes}
\Output{Updated reputation scores of $q$'s $k$-bucket contacts
 }
 \BlankLine
 \If{$u$ is in $\visited$ list}{return\;}
 \BlankLine
 \If{$u$ = $q$}{return\;}
 \BlankLine
 $\increment$ by $1$ the reputation score of $k$-bucket contact $u$ of querying node $q$\;
 $\amark$ node $u$ as $\visited$;
 \BlankLine
 \For{each node $v$ with an edge from $u$ to $v$ in $G$}{    
 	\KwFn{$q$, $G$, $v$}\;
 } 
\caption{\sc{Updating reputation score for Kad-\sys}}
\label{algo:update-rep-kad}
\end{algorithm}

\emph{Using the reputation scores.} 
During the lookup process, the querying node uses its local reputation
information to pick the best $\alpha$ nodes from the appropriate
$k$-bucket.
Each of the queried nodes in turn collaborates by providing the best
$\beta$ contacts from the appropriate $k$-bucket using their reputation
scores from first-hand observations. Thus, at each step the basic Kad
algorithm of cutting the remaining distance in half is honored, except
that a better choice is made while picking nodes from within the
$k$-buckets. We also modified the bucket eviction policy in which
regular Kad replaces the least-seen node when a new node is being added
to an already full $k$-bucket.
\ksys instead replaces the node with the lowest reputation score,
breaking ties by replacing the least seen of the least reputed nodes.


\paragraphX{Attack model under Kad-\sys.} The $k$-bucket population can
change dynamically in Kad as nodes are encountered during the lookup
process. To model the most effective attack possible, we make the
following change to the attack model for Kad. When the attacker chooses
not to attack a query, he attempts to provide the correct closest
replica root at the end of the lookup to maximize his reputation
scores. In addition, he attempts to pollute routing tables with
malicious nodes during the lookup process. In particular, a malicious
node knows about all of the other malicious nodes and can provide them
as answers to a query. He does so only as long as they are at least one
bit closer in the $\xor$ distance to the target, since more distant
malicious nodes will be ignored. This approach allows the attacker nodes
to be seen and possibly selected for being added to $k$-buckets more
than would be expected. At the same time, malicious nodes can still
provide a correct lookup result upon termination of the lookup process,
ensuring a positive effect on reputation.


%

\subsection{Handling Churn} \label{sec:reds:dynamic}


In our simulations (see Section~\ref{sec:eval}) we evaluate \sys under
a dynamic network with churn. In large P2P systems peers
generally leave and rejoin the system at irregular intervals. This churn
makes relying on predictions based on past behavior inaccurate at larger
time scales. The attacker can also modulate the behavior of his peers to
manipulate the reputation system. Several techniques can be used to
mitigate this risk.

We explored techniques such as exponentially weighted moving
  average (EWMA) to cope with churn, but, as we show in
Section~\ref{sec:eval-churn}, it is effective to update the scores with
equal weight to older and newer results. EWMA is still recommended to
deal with oscillation attacks, as described in
Section~\ref{sec:security:oscillation}.
We also considered various exploitation-versus-exploration tradeoffs,
but deterministically picking the node with the highest A-Boost score
provided the best results. Since the finger and its predecessor have the
same initial reputation scores, one of them will be picked at random at
the beginning. If the selected node loses reputation because of failed
searches, then the other node is picked. Eventually, each node is explored if
one or both display failures, and thus we found both exploitation and
exploration taking place on an as-needed basis. Our findings for
exploration validate analysis from Section~\ref{sec:security}. 

\subsection{Shared reputation scores}\label{sec:reds:sReDS}

An intuitive idea for improving ReDS is for peers to share reputation
information with each other. Shared reputation is beneficial for and
even a central part of many reputation systems in the context of
free-rider prevention~\cite{hoffman09survey}. We thus explore how shared
reputation could work in ReDS and how well it would work.

Unlike reputation systems in many contexts, ReDS peers cannot make use
of reputation information shared by arbitrarily selected peers. They can
only use reputation from nodes who share the same fingers.
We thus aim to identify and maintain a list of the nodes with shared
fingers and regularly share reputation information with them.

Specifically, we worked out a shared reputation scheme for Halo, which
has deterministic finger selection. We expect that shared reputation in
Kad will be significantly harder, since $k$-buckets are populated
opportunistically. As we show in our experiments and analysis, shared
reputation is ineffective in the face of malicious reputation sharing,
and thus we do not attempt to devise a scheme for Kad.
In the context of Halo we can define two nodes that share the same
finger $f$ to be \emph{joint knuckles} of each other for finger $f$. In
this approach each node maintains a list of joint knuckles for each of
its fingers. The list can be maintained by periodically performing the
knuckle search on each finger, which is already a Halo primitive. The
node then incorporates the scores of these joint knuckles into a score
for its finger. We divide the scheme into two phases: (I) sharing
reputation scores with joint knuckles and (II) calculating the shared
reputation scores.

%

\paragraphX{Phase I: Sharing.} We divide time into a series of epochs
based on the assumption of loosely synchronized clocks (e.g. with
NTP). At the beginning of each epoch $t_i$, a node compares
its first-hand reputation score with
its score from epoch $t_{i-1}$. If the score has changed, then it
broadcasts the updated score to its joint knuckles.


%

\paragraphX{Phase II: Calculating reputation.} After receiving all
updated scores,
the node can calculate the shared reputation score for each of its
fingers. Taking the average reputation score is not robust to
self-promotion and slandering attacks. Wagner goes into great detail on
aggregation methods that are suitable for security applications, finding
that median is a strong solution~\cite{wagner04agg}. We note, however,
that median {\em always} fails when the number of attackers is more than
50\%. Having so many attackers among one's joint knuckles is possible
due to the small sample size. Since we have our own reputation
information collected from local observations, we can do better.

We now describe a novel scheme for reputation aggregation called {\em
  Drop-off}, in which scores close to the node's own local scores are
more likely to be considered for a final aggregation step. The key
assumption in this approach is that the score from first-hand
observations
is a better approximation of the correct score than scores from
slandering or self-promoting attackers. We aim to balance between
accepting and hopefully gaining from others' reputation information
(which may be different from our own) while trying to limit
vulnerability from slandering and self-promotion attacks. 

Let $r_k(f)$ be the first-hand reputation score of finger $f$ as
measured by knuckle $k$. $k$ receives a reputation score for $f$ from
joint knuckle $j$, which is $r_j(f)$. $k$ then calculates $w =
1-|r_j(f)-r_k(f)|$ and places $r_j(f)$ into a {\em scoring bin} for $f$
with probability $w$. Intuitively, the further $j$'s score is from
$k$'s, the less likely it is to be included in the scoring bin. $k$'s
shared reputation score for $j$ for the current epoch is the median of
the scores in the scoring bin.

\paragraphX{Slandering and Self-Promotion.}
%
Slandering and self-promotion attacks are the most prominent attacks
against a shared reputation system, aiming to make a targeted peer
select malicious fingers for lookup operations. To show the value of
\sharedrep, we must show its resilience to these attacks.


%
%

We now derive an equation to estimate the expected score that the
Drop-off method would provide. Let $k$ be the knukle of a finger $f$ and
let $r_k(f)$ be $k$'s reputation score for finger $f$ based on
first-hand observations. For both slandering and self promotion attacks,
let us assume that $r_m(f)$ is the reputation score of $f$, received
from the malicious knuckles and $r_h(f)$ is the reputation scores of
$f$, received from honest knuckles. Malicious scores are assumed to all
be the same, as are honest scores, for simplicity of analysis. Let
$n_{h}$ be the number of $k$'s joint knuckles of $f$ that are honest and
$n_m$ be the number that are malicious. Finally, let $d_h = |r_h(f) -
r_k(f)|$ and $d_m = |r_m(f) - r_k(f)|$ be the differences between $k$'s
score and the scores from its honest and malicious joint knuckles,
respectively. Based on the Drop-off algorithm, $(1-d_h)$ is the
probability of selecting the score from an honest knuckle to calculate
the median of the shared scores. 
%
Letting $p$ be the probability that the number of honest
nodes selected is more than the number malicious nodes selected, we
have:

{\small
\begin{equation*}
p=\sum_{i=1}^{n_h}\sum_{j=0}^{i-1}\binom{n_h}{i}{{(1-d_h)}^{i} d_h^{n_h-i} \binom{n_m}{j}{(1-d_m)}^{j} d_m^{n_m-j}}
\end{equation*}
}
 
Let $q$ be the probability that the number of malicious nodes and honest
nodes selected are the same, meaning that the median will be calculated
as $\frac{r_h(f)+r_m(f)}{2}$. Assuming that at least one honest node and
malicious node are selected, and letting $n'=min(n_m, n_h)$, we get:

\begingroup
\setlength\abovedisplayskip{0pt}
\[
q=\sum_{i=1}^{n'}\binom{n_m}{i} {(1-d_m)}^{i} d_m^{n_m-i} \binom{n_h}{i} {(1-d_h)}^{i} d_h^{n_h-i}
\]
\endgroup

In total, the expected Drop-off score $E[s_{\delta}]$ is given by:
\begin{equation*}
E[s_{\delta}] = p r_h(f) + q \frac{r_h(f)+r_m(f)}{2} + (1-p-q) r_m(f).
\end{equation*}

To illustrate the effect of the \sharedrep approach, let us consider the
following simple numerical example of a self-promotion attack against a
knuckle $k$. Suppose that for calculating the score of a malicious
finger $f$, $k$ finds 11 joint knuckles, of which six are attackers. Let
us say that the ``true'' reputation score for $f$ is 0.1 (only 10\% of
the searches through it will succeed), while $k$'s estimated score from
first-hand observations is currently 0.3.

Assume for simplicity that all six attackers claim that their reputation
score for $f$ is 1.0, while all five honest nodes report a score of 0.1
for $f$. The average score is 0.59, which is much higher than the true
score. The median has reached the {\em breakdown point}, since more than
half of the nodes are malicious; the median score is 1.0. For
\sharedrep, we first must examine the population of the bucket. The
expected number of honest nodes in the bucket is four, while the
expected number of malicious nodes is 1.8. In 88\% of the cases, the
honest nodes form a majority of the bucket and the \sharedrep score is
0.1. The overall expected \sharedrep score is 0.17.



The system works similarly against slandering attacks. With this
approach, the \sharedrep scheme provides much better scores than taking
the average. It also provides a way to avoid the breakdown point that
the median faces against a majority of attackers as joint knuckles.
The simulation results presented in Section~\ref{sec:results-shared}
bear out our analytical findings that \sharedrep is a novel improvement
in calculating shared reputation scores over other techniques. However,
they also show only limited benefits of sharing in ReDS. Further, we
present an attack on shared reputation for ReDS in
Section~\ref{sec:shared-use}.

%



\section{Experimental Evaluation}
\label{sec:eval}

We now present results from extensive simulations of \hsys and \ksys. We
first describe our experimental setup, and then we present the simulation
results.
\subsection{Experimental setup}
\label{sec:setup}
We built simulators for both \hsys and \ksys in Java. Each simulator
includes the basic lookup mechanism of the network,\footnote{We use  ``network'' to indicate the underlying DHT,
  i.e. Chord or Kad.} the A-Boost reputation tree for each node,
collaborative boosting, a model for node churn, and attacker models
specific to the network (see Section~\ref{sec:model:attack}).


\paragraphX{Setup for \hsys.}
All our simulations for \hsys were run for networks with 1000
nodes.\footnote{Larger networks can be simulated, but take an
  impractically long time to finish since each node in the network must
  build up reputation information through lookups.} In our experiments
we use a redundancy of 10 as suggested for regular Halo with 1000 nodes
in the network. 

\paragraphX{Setup for \ksys.}  
Most of our simulations for \ksys were run for networks with $10,000$
nodes. The largest simulation was run for 100,000 nodes. In Kad and
\ksys, we initialize the system with $n_l$ lookups per node to populate
the $k$-buckets. The difference between collaborative boosting and
A-boost is in the selection process of the $\beta$ nodes returned by an
intermediate queried node during every step of a lookup process, as
described in Section~\ref{sec:kad-reds}.
Also as discussed in Section~\ref{sec:kad-reds}, we impose an attacker
model on Kad that includes routing table pollution, since routing tables
are populated dynamically, and we modify the bucket replacement policy
to replace low reputation nodes.
Kad has inherent redundancy controlled by
parameters $k$ and $\alpha$ as described in Section~\ref{kad}. We
used $k$=10 and $\alpha$=7 redundancy for most of the simulations.

\paragraphX{Churn.} To evaluate how the network handles node churn, we
add and remove nodes probabilistically after each lookup (which are
treated as atomic operations). The probability of a given node joining
or leaving after a given lookup is set based on the intended churn rate
for that simulation run. For example, in a simulation with $n = 1000$
nodes, a colluding fraction of $c = 20\%$, $l=250$ training lookups, and
a churn rate of $r = 25\%$ over the whole simulation (i.e., in a network
with 1000 nodes, on average 250 nodes leave the network and 250 new
nodes join the network over the course of the simulation),  the
probabilities for a single node are $p_{leave} = p_{join} = 0.00125$,
calculated as
\begin{equation*}
p = \frac{1}{\frac{l \cdot (1 - c) \cdot n}{r \cdot n}} = 
 \frac{r}{l \cdot (1-c)}
\end{equation*}
where the $(1-c)$ stems from the fact that only honest nodes do training
lookups.


\paragraphX{Nodes chosen for lookups.} For all simulations, all honest
nodes are selected in a random permutation as querying nodes. For A-Boost
and collaborative boosting, this helps to build the reputation trees of
all the honest nodes. For all \sys simulations, nodes use
\emph{deterministic maximum score} as described in
Section~\ref{sec:reds:dynamic} to select fingers for routing.

\paragraphX{Shared reputation.} When shared reputation is used, all
honest nodes are trained and queried similarly, with the addition that
nodes gather shared reputation information from joint knuckles when
evaluating fingers. Shared reputation is only available in \hsys, as
described in Section~\ref{sec:reds:sReDS}.

\paragraphX{Sampling.}  For some of the results in this section, we used
a {\em continuous simulation mode}, in which the network is sampled at
regular intervals as the simulation progresses. This allowed us to
monitor the evolution of the failure rate as nodes learn more
information about the network and as nodes join and leave the
network. To achieve this, we conduct alternating phases of $n_t$ {\em
  training lookups}, during which reputation scores are set, and $n_l$
{\em probing lookups}, during which the failure rate of lookups was
recorded. Probing can be thought of as taking a snapshot of the state of
the network. One set of training lookups and probing lookups is a
\emph{slot}.
We then took the failure rate achieved in the steady-state as the final
result. Since continuous simulations show changes over time, they
represent a single (often very long) simulation run.

In other (non-continuous) simulations, we simply run a long training
phase and then a single probing phase at the end. Each data point in
these graphs corresponds to an average value with standard error bars
from $n_i$ different instantiations of the DHT, where we typically set
$n_i=10$.



\subsection{Long-term performance under churn}
\label{sec:eval-churn}

\begin{figure}[t]
\centering
   \includegraphics[width=\scalefac\columnwidth]{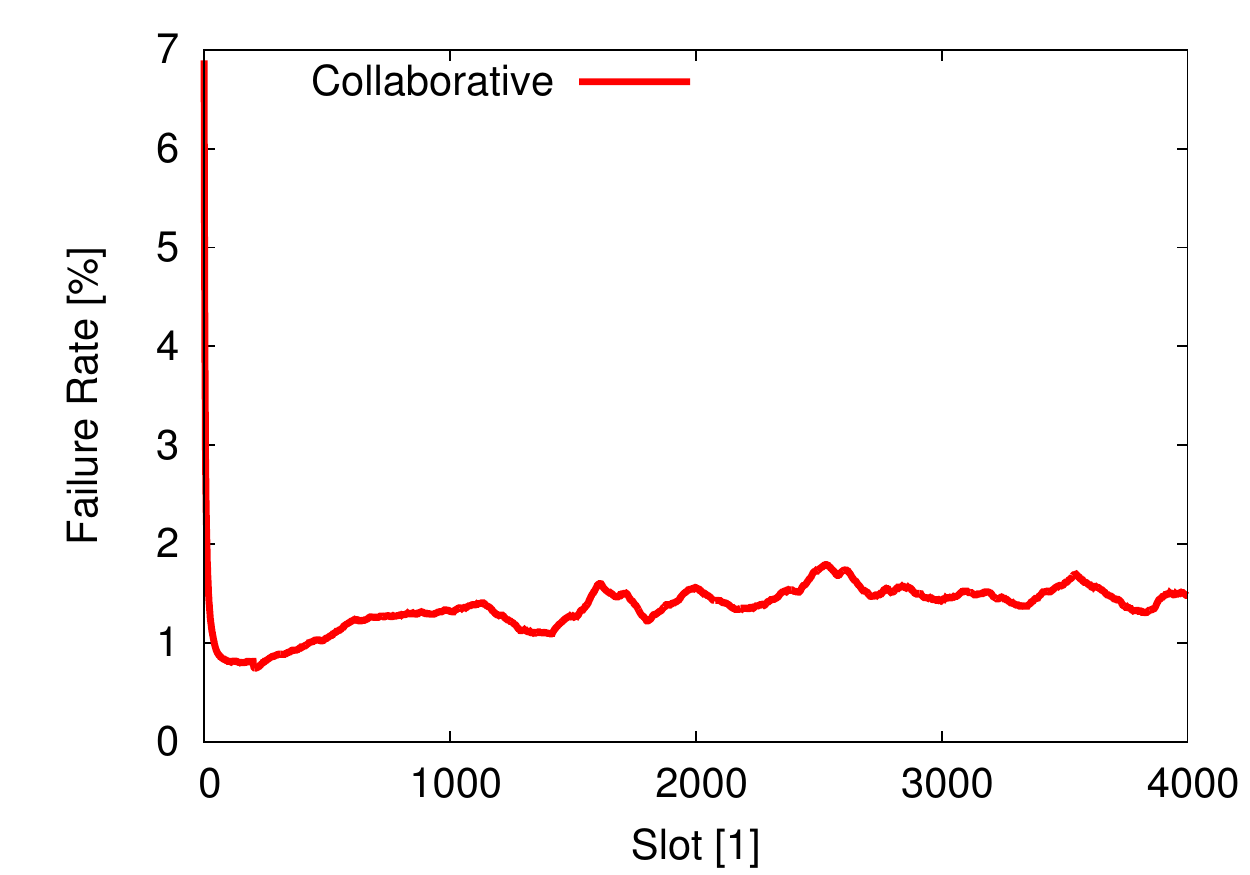}
 \caption{{\bf Halo-ReDS (Continuous).} The evolution of the failure
   rate. The failure rate appears to stabilize around a steady-state
   value.}    \label{fig:churn_c02}
\end{figure}

One issue with reputation in a DHT is that as nodes join and leave the
network (i.e., under node churn), reputation information becomes
stale. 
We also seek to determine whether \sys performance reaches a reliable
steady state as churn continues to affect the system.

To these ends, we ran \hsys experiments in continuous simulation mode in
which all nodes were replaced on average once every 800 lookups (160
slots), i.e., by the time a node has done 800 lookups, on average all
nodes in the network have been replaced once. We then let this
simulation continue for altogether 20,000 lookups per node (4,000 slots,
20 million total lookups), and calculate the average failure rate over
the latter half of the simulation (i.e., the latter 10,000 lookups or
2,000 slots) to get a \emph{steady state} value for the failure rate.


Figure~\ref{fig:churn_c02} shows the results for $c=20\%$ and attack
rate $a=1.0$.  The failure rate quickly drops from a relatively high
rate of $7\%$ early on, when no reputation information is available, to
less than $1\%$. As churn sets in, the failure rate increases until it
reaches a steady state of around $1.5\%$, indicating that removing
reputation information for fingers that leave the network to a large
extent solves the problem of stale reputation, limiting the effect of
node churn on \sys.


\begin{figure}[t]
  \centering \subfigure[Attack rate $a=1.0$. A-Boost is unable to
    improve over regular Halo due to node churn. Collaborative boosting,
    however, performs significantly better throughout.]{
    \label{fig:failure_rate_a1.0}
  \includegraphics[width=\scalefac\columnwidth]{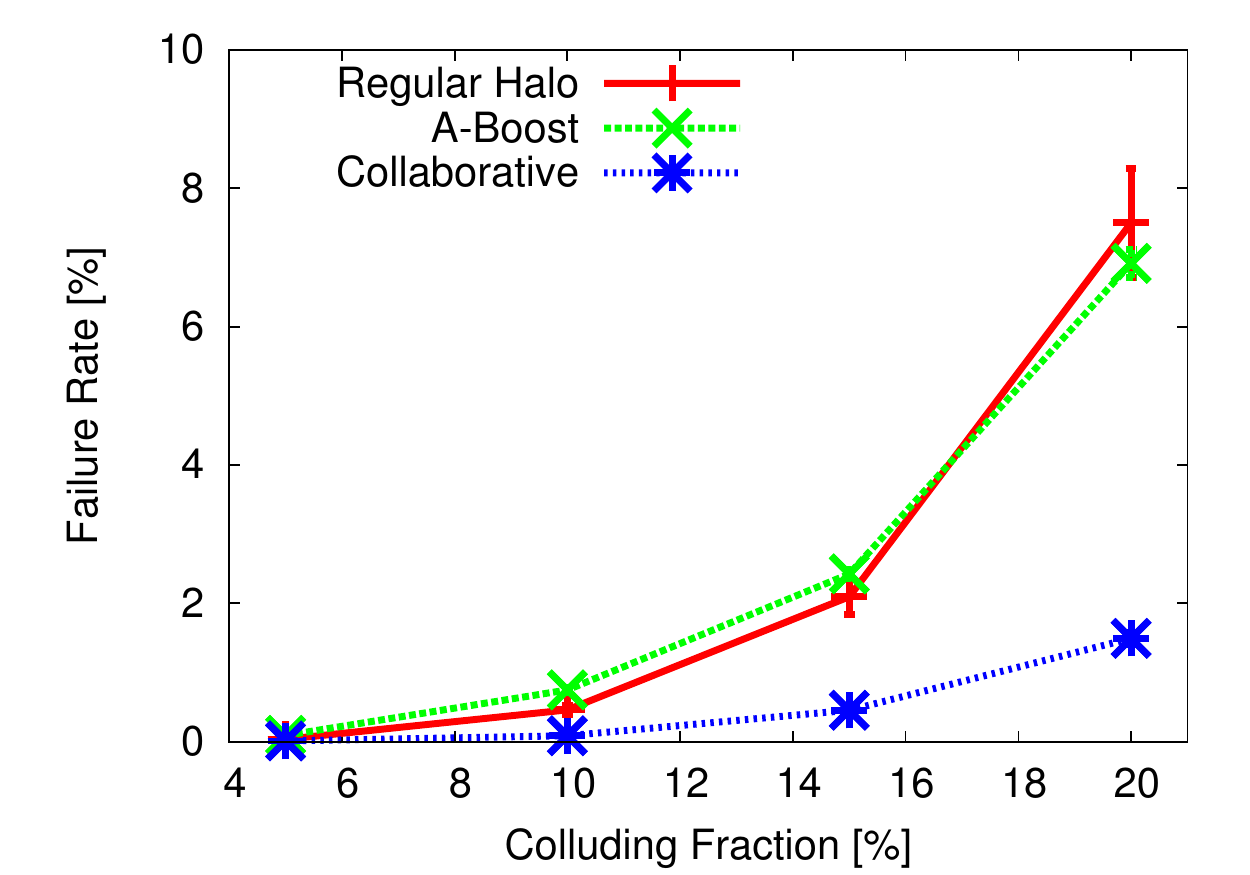}}
  \subfigure[Attack rate $a=0.5$. A-Boost now performs worse than
    regular Halo, while collaborative boosting again performs
    significantly better than either regular Halo or A-Boost.]{
    \label{fig:failure_rate_0.5}
    \includegraphics[width=\scalefac\columnwidth]{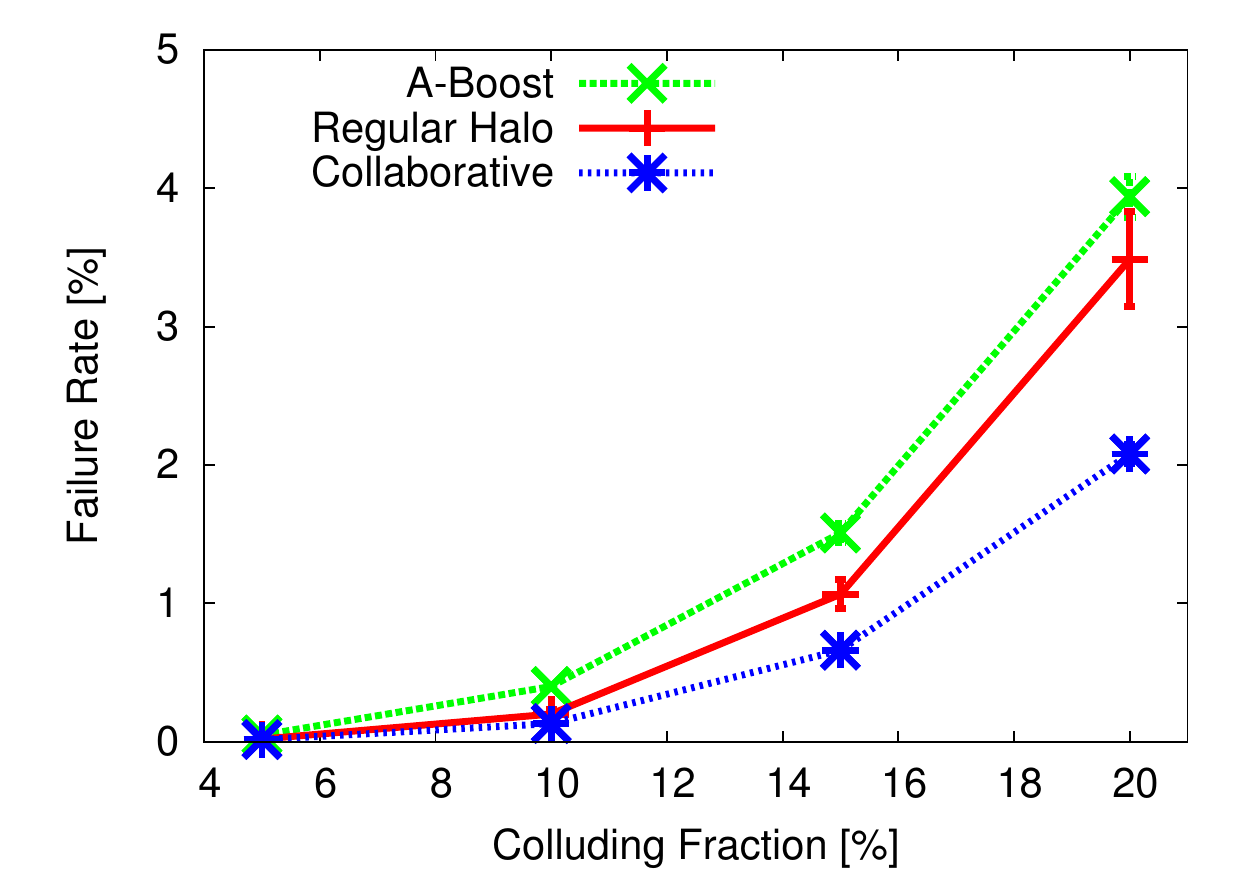}}
  \caption{{\bf Halo.} These graphs compare the failure rates
    for different Halo algorithms and colluding fractions.}
  \label{fig:failure_rate}
\end{figure}

\subsection{Comparison of different Halo schemes.}
\label{sec:eval-schemes}

We now compare the failure rates of different schemes (regular Halo,
A-Boost, and collaborative boosting) for different colluding rates using
non-continuous simulation runs. Figure~\ref{fig:failure_rate_a1.0} shows
the failure rate for regular Halo, A-Boost, and collaborative boosting
for an attack rate of $a=1.0$. Due to node churn, A-Boost performs
roughly the same as regular Halo.
Collaborative boosting, on the other hand, significantly reduces the
failure rate, down to 1.5\% for a colluding fraction of 20\%, an
improvement of around 79\% over regular Halo and 73\% over A-Boost.

The results for an attack rate of $a=0.5$ are shown in
Figure~\ref{fig:failure_rate_0.5}. Note that in such a scenario, there
is only half as much information available to the reputation system
about attackers. The failure rate of regular Halo is now approximately
half of what it was for an attack rate of $a=1.0$, because only half of
all lookups are attacked. A-Boost is now performing worse than regular
Halo. Collaborative boosting, on the other hand, is again performing
much better than regular Halo and A-Boost, reducing the failure rate by
40\% and 50\% compared to regular Halo and A-Boost, respectively.



A-Boost suffers under churn because it cannot easily adapt to changes
beyond its fingers. The reputation trees of nodes that did not see the
change in their fingers' fingers (and further down the tree) do not
account for these churn events.
In collaborative boosting, however, nodes can rely on their fingers to
adapt to changes further down the lookup paths. This greatly improves
the speed at which lookup paths are modified to address churn events.
%

\subsection{Performance of \ksys.}
\label{sec:eval-kad-perf}
%

\begin{figure}[t]
 \centering 
  \includegraphics[width=\scalefac\columnwidth]{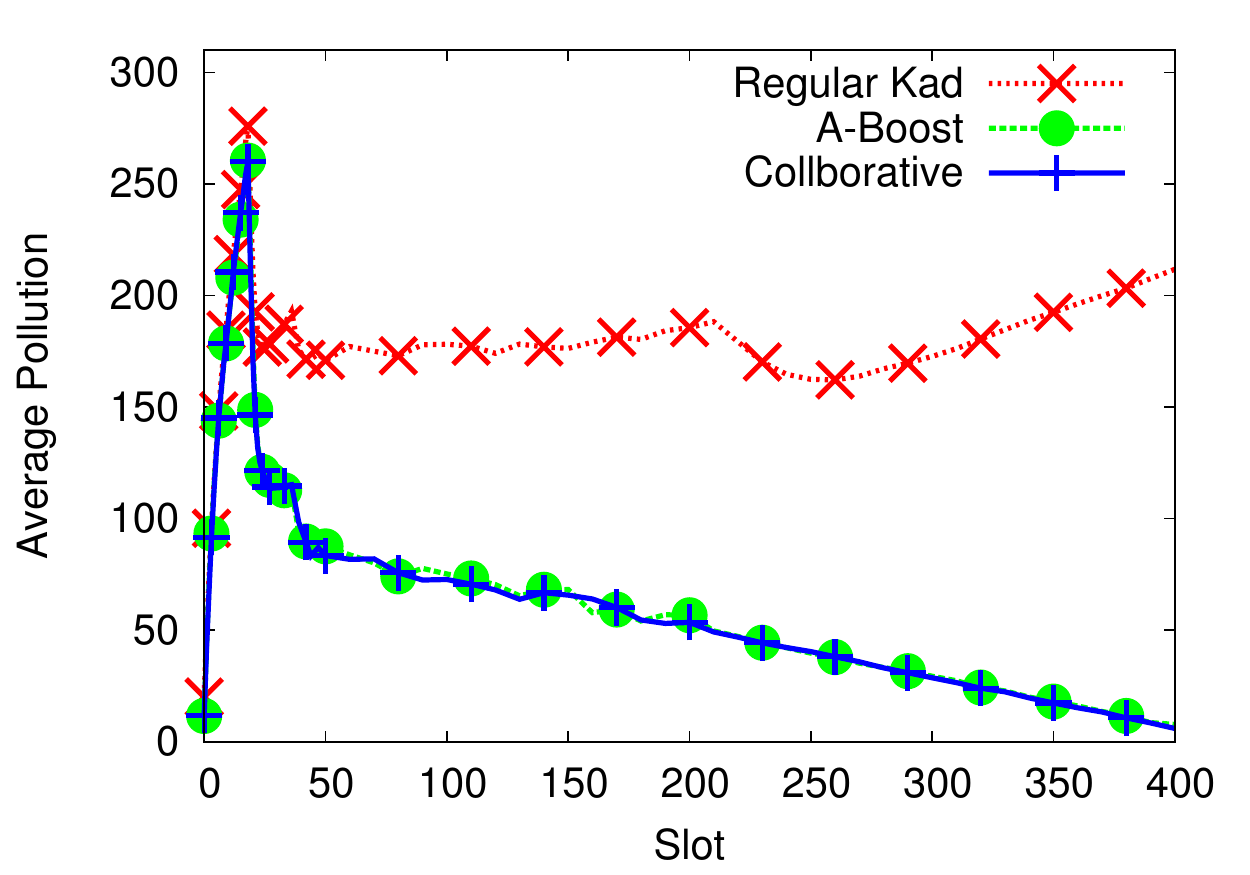} 
\caption{{\bf Kad (Continuous).} Evolution of routing table pollution
  over the lifetime of an experiment with $c=20\%$ and $a=1.0$. Regular
  Kad faces increasing pollution over time, while \ksys identifies and
  removes malicious fingers from $k$-buckets.}
\label{fig:kad_pollution}
\end{figure}

We now turn towards evaluating our design for \ksys. Our primary
concerns in \ksys are whether routing table pollution can be overcome,
the general effectiveness of the design under different redundancy
parameters, and the performance of A-Boost and collaborative boosting
under churn.

\paragraphX{Routing Table Pollution.} We find that routing table
pollution is the most critical factor in Kad and \ksys
performance. Thus, we first seek to understand the extent of routing
table pollution these systems. To this end, we perform a continuous time
simulation with 10,000 nodes under $r=25\%$ churn. Each of the nodes
performs 100 lookups in order to populate the $k$-buckets and build the
reputation system. We divide the simulation time into 400 slots of 2500
lookups each.

As discussed in Section~\ref{sec:kad-reds}, attacker nodes are
attempting to get their own nodes into as many routing tables as
possible. In \ksys, however, attacker nodes with low reputation scores
will have little chance to be selected as the next hop contacts in
future lookups and will eventually be kicked out of many
$k$-buckets. Figure~\ref{fig:kad_pollution} shows the pollution of
routing tables over the training period for both regular Kad and
\ksys. We see that the reputation system is very effective, leading to a
decreasing rate of pollution of routing tables with \ksys, compared with
increasing pollution rates in Kad.

\begin{figure}[t!]
\centering
  \subfigure[Kad.]{
    \label{fig:kad_f_areg}
    \includegraphics[width=\scalefac\columnwidth]{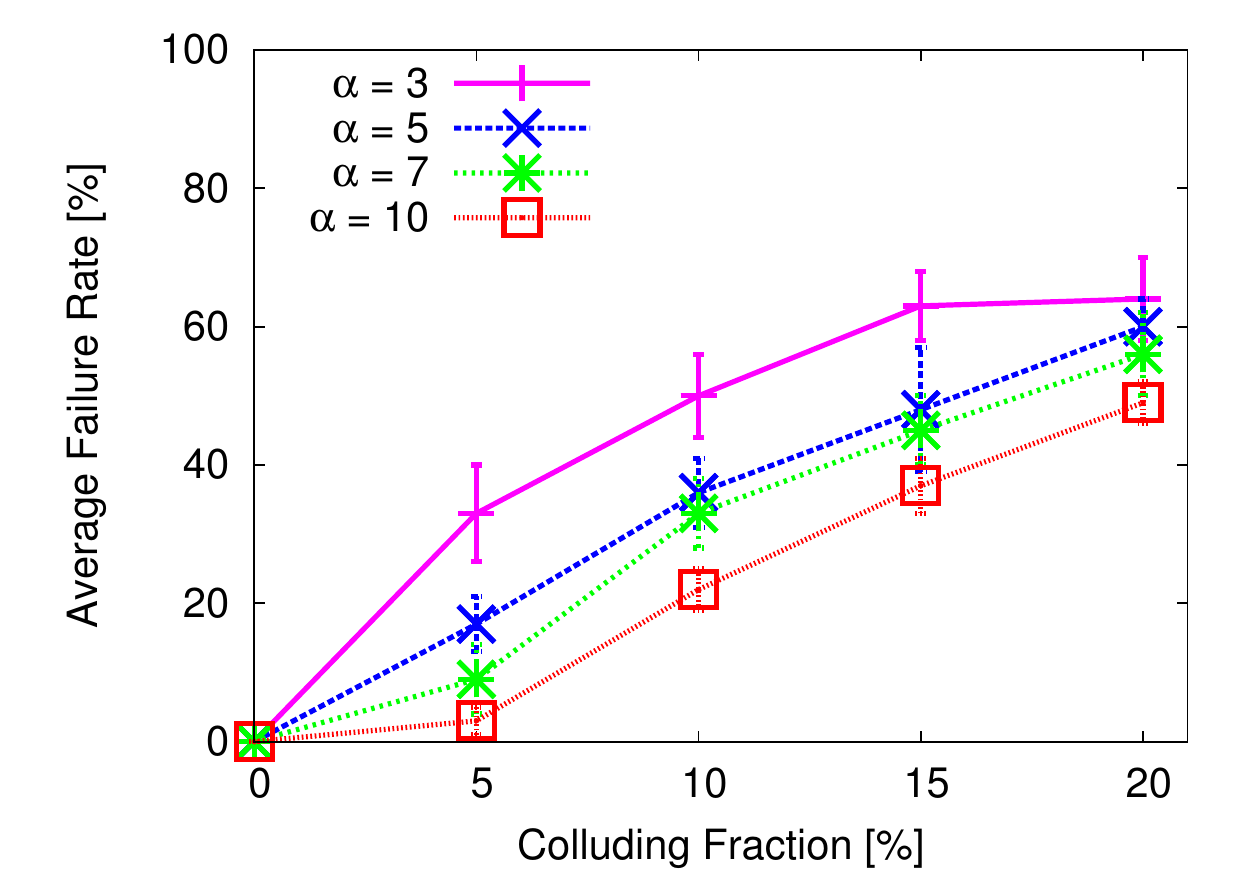}}
  \subfigure[\ksys, collaborative boosting. Note that the $y$-axis is
    from 0-45\%]{
    \label{fig:kad_f_a}
    \includegraphics[width=\scalefac\columnwidth]{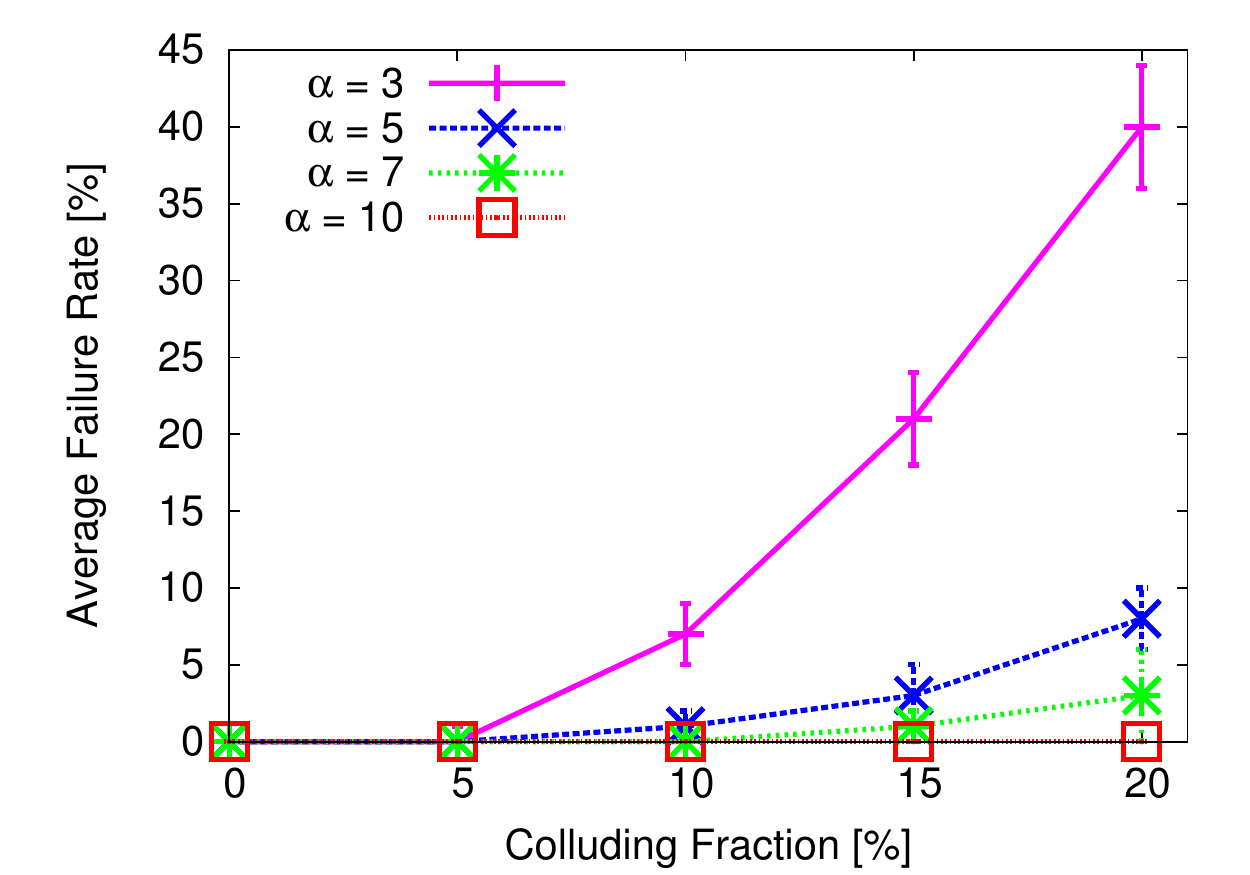}}
 \caption{{\bf Kad (No Churn).} Failure rates with $k=10$, $a=1.0$, and
   varying $\alpha$ and $c$. Higher redundancy ($\alpha$) improves both
   systems, but \ksys is much better than Kad for $\alpha > 3$.}
      \label{fig:kad_f_graphs}
\end{figure}

\begin{figure}[t!]
\centering
  \subfigure[Attack rate of $a=1.0$.]{
    \label{fig:kad_churn_a100}
    \includegraphics[width=\scalefac\columnwidth]{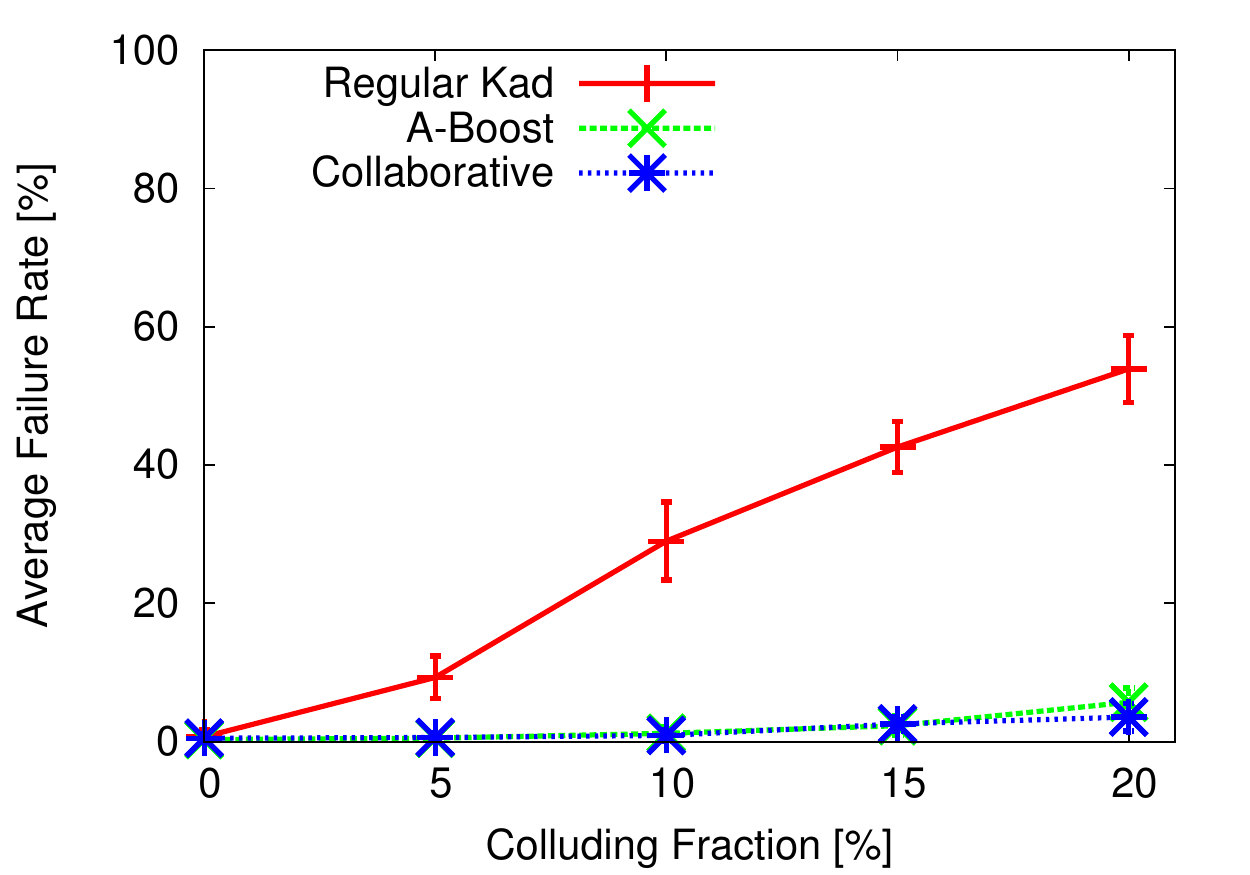}}
 \subfigure[Attack rate of $a=0.0$.]{
    \label{fig:kad_churn_a0}
    \includegraphics[width=\scalefac\columnwidth]{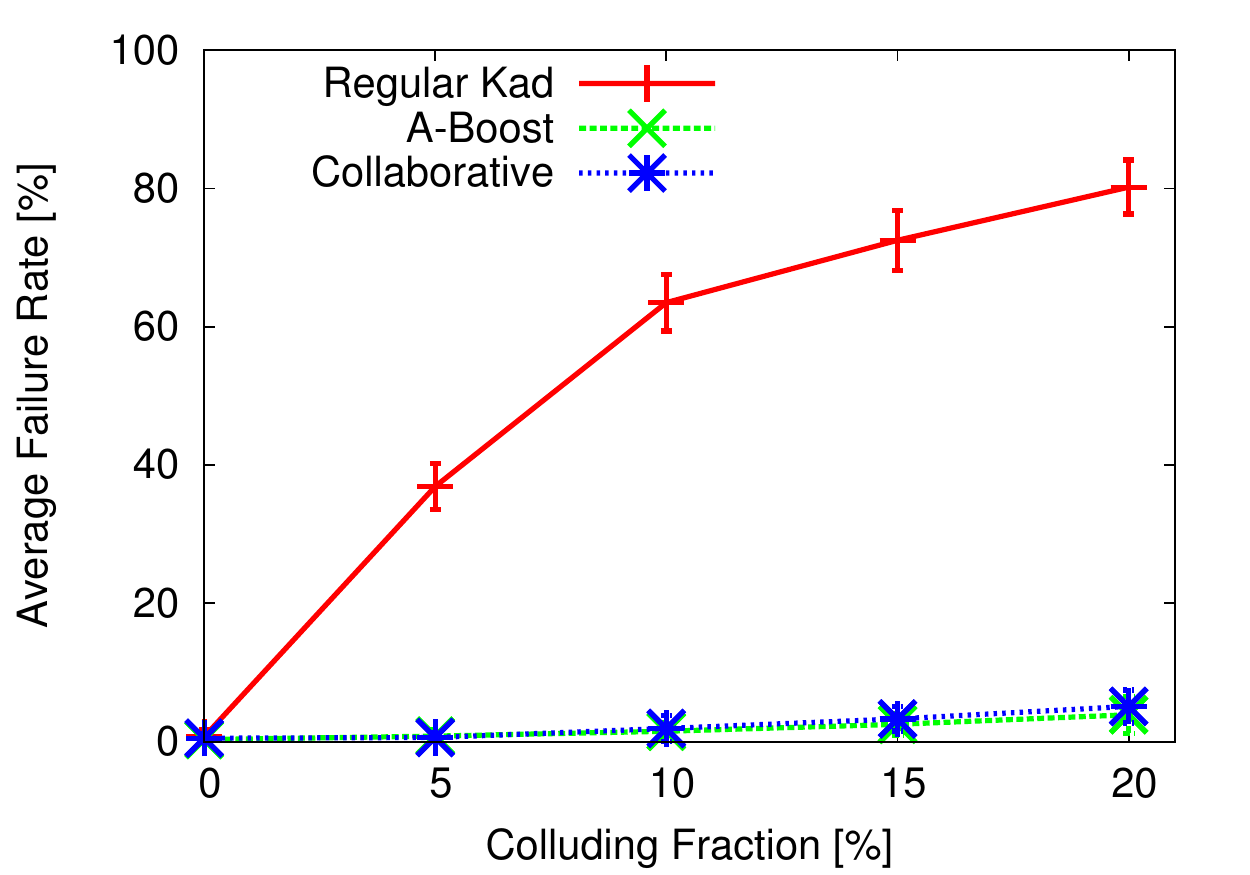}}
 \caption{{\bf Kad.} Failure rates of different \ksys schemes for
   different colluding fractions $c$. Both A-Boost and Collaborative are
   much more effective than Kad.}
      \label{fig:kad_churn_graphs}
\end{figure}


\paragraphX{Redundancy.} We also explore the performance of \ksys in
extensive non-continuous simulations. First, we break down the
performance in detail without churn. Figure~\ref{fig:kad_f_graphs} shows
the effect of redundancy on regular Kad without churn.
We use an attack rate of $a=1.0$. Figure~\ref{fig:kad_f_areg} shows how
the failure rate decreases as the redundancy increases. However, even
with a high redundancy of $\alpha=10$, the failure rate when $c=10\%$ is
over 21\%. Collaborative boosting dramatically improves the
system. Figure~\ref{fig:kad_f_a} shows failure rates for \ksys. With a
lower redundancy of $k=10$ and $\alpha=5$, the failure rate when
$c=10\%$ is less than 1\%. When $c=20\%$, $k=10$, and $\alpha=7$, Kad
has a $56\%$ failure rate, while \ksys has a $3\%$ failure rate, a 95\%
decrease.


\paragraphX{Comparison Under Churn.} We now compare the performance of
Kad and both collaborative and A-boost versions of \ksys under churn. 
We present results for $r=25\%$ churn, and $k=10$ and $\alpha=7$
redundancy. Figure~\ref{fig:kad_churn_a100} shows the failure rate of
regular Kad, A-Boost, and collaborative boosting for an attack rate of
$a=1.0$. Performances of A-Boost and collaborative are almost the same
(within the margin of error), significantly reducing the failure rate
of Kad down to 4-5\% from 54\% for $c=20\%$. By comparing
Figure~\ref{fig:kad_f_a} and Figure~\ref{fig:kad_churn_a100} we note
that, the performance of Kad-\sys with $k$=10 and $\alpha$=7 degrades
from 2-3\% failure rate to 4-5\% due to the 25\% churn. 

The results for an attack ratio of $a=0.0$ are shown in
Figure~\ref{fig:kad_churn_a0}. In our attack model
(Section~\ref{sec:kad-reds}), the attackers are always trying to
pollute the routing tables even when they are not attacking.
Figure~\ref{fig:kad_churn_a0} shows that failure rate slightly increases
for 0\% attack rate compared to the failure rate for 100\% attack
rate. We will discuss more about our attack effectiveness in
Section~\ref{sec:overall_effectiveness}.

Over all scenarios we tested, we find that performance is improved by at
least 93.4\% with \ksys compared to regular Kad.

\subsection{Overall attack effectiveness.}
\label{sec:overall_effectiveness}
In this experiment, we show the \emph{overall attack effectiveness} when
honest nodes use 
collaborative boosting. The overall attack effectiveness is the maximum
continuous failure rate that the attacker can achieve when his nodes use
a consistent attack rate, i.e. without jumping to 100\% attack rate for
evaluation. This allows us to identify the best that the attacker could
do consistently over time.

\paragraphX{Halo.} We first performed the experiment for both regular
Halo and A-Boost. In regular Halo, the attack effectiveness grows
linearly with the attack rate as expected. In A-Boost, which is
generally about as effective as regular Halo under churn, the results
are quite similar. For $a=0.1$ attack rate, both systems had less than
1\% failure rate for $c=$5\% to 20\%. For $a=0.5$ and $c=$20\%, the
failure rate is between 3.5\% to 4.0\%, and for $a=1.0$ and $c=$20\%,
the failure rate is between 6.9\% to 7.5\%.

\begin{figure}[t!]
  \centering 
  \includegraphics[width=\scalefac\columnwidth]{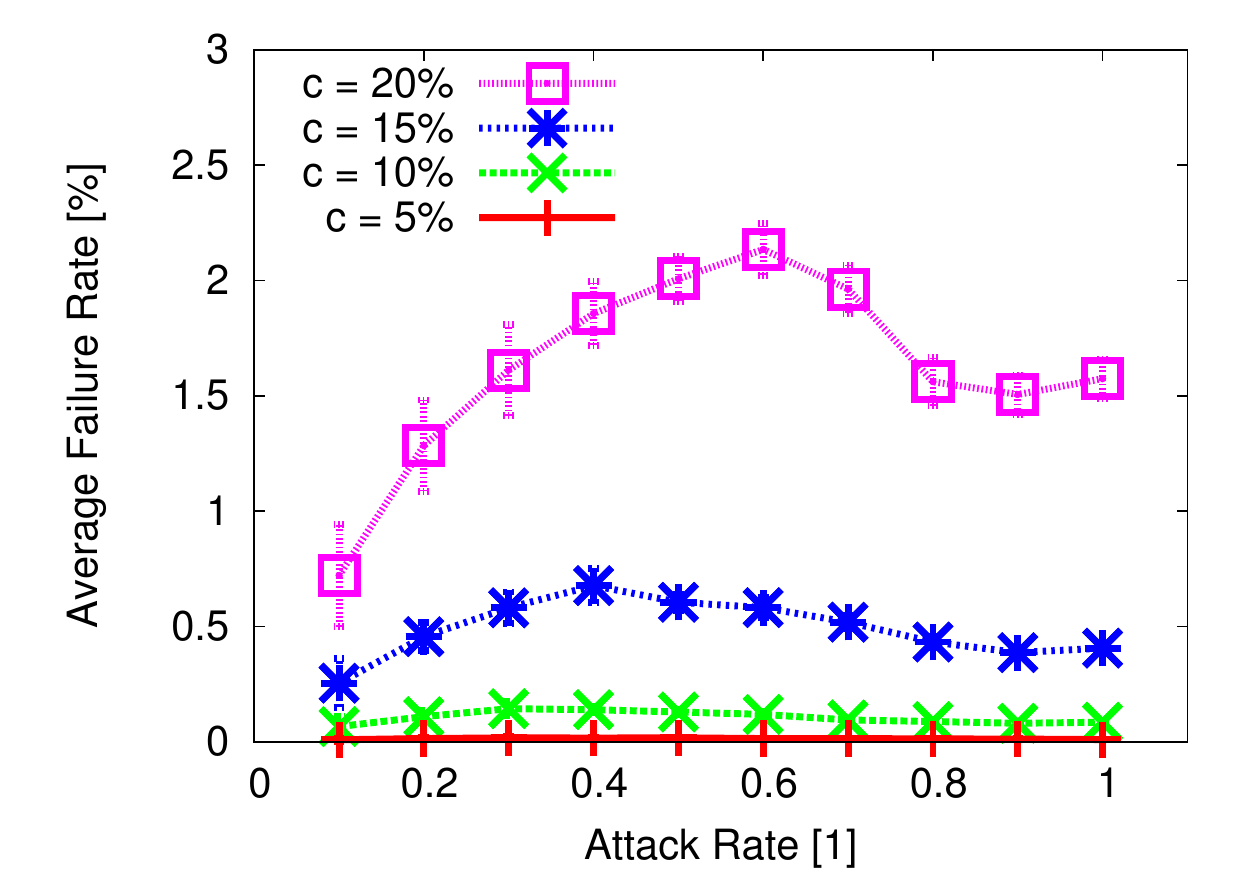}
  \caption{{\bf \hsys.} The failure rate for different continuous attack
    rates, with $c=20\%$. Attacker effectiveness peaks at less than
    $a=1.0$.}
  \label{fig:overall_collaborative} 
\end{figure}

Figure~\ref{fig:overall_collaborative} shows the overall attack
effectiveness that the attacker can achieve against collaborative
boosting. We see that increasing the attack rate up to a point results
in more lookup failures. Beyond a certain attack rate, e.g., at 60\% for
a colluding fraction of $c=20\%$, the overall failure rate goes back
down. Specifically, for $c=20\%$, the effectiveness peaks at a $2.1\%$
failure rate, for $c=15\%$ at $0.7\%$, for $c = 10\%$ at $0.014\%$, and
for $c = 5\%$ at a tiny fraction of a percent. Comparing the overall
effectiveness of collaborative boosting to A-Boost and regular Halo,
\sys reduces effective failure rates by up to 70\% for $c = 5\%, 10\%$,
and by up to 80\% for $c = 15\%, 20\%$.
The reason for the peak in effectiveness is that with lower attack
rates, fewer lookups are being subverted, while with higher attack
rates, the malicious nodes are more easily detected by the reputation
system and are no longer used.

Despite the ability of attackers to operate at a peak rate, we note that
for colluding fractions of 20\% and below, no matter what rate the
attackers attack with, \emph{collaborative boosting limits their
  effectiveness to below 2.1\%}.

\begin{figure}[t!]
\centering
  \subfigure[Kad.]{
    \label{fig:kad_churn_areg}
    \includegraphics[width=\scalefac\columnwidth]{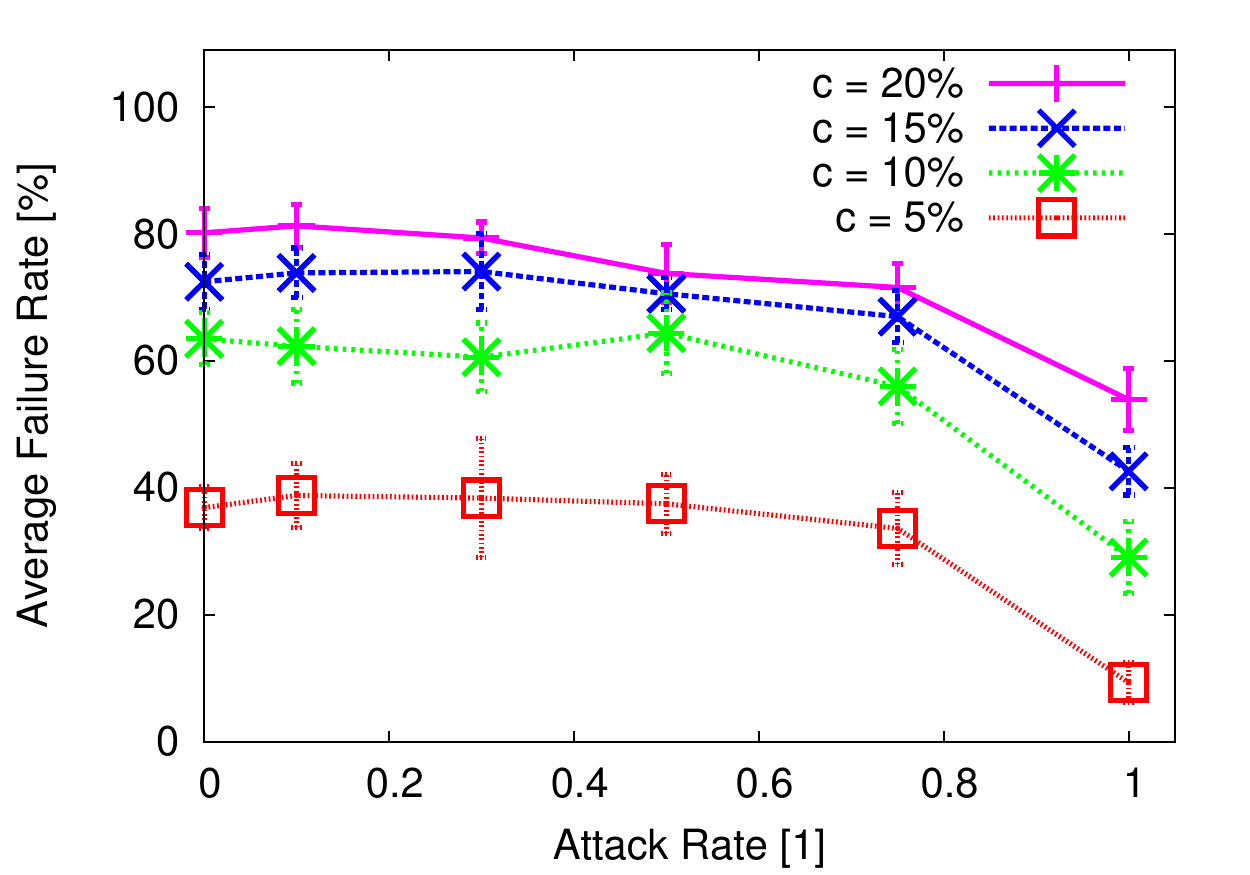}}
 \subfigure[\ksys. $y$-axis ranges from 0-8\%.]{
    \label{fig:kad_churn_a}
    \includegraphics[width=\scalefac\columnwidth]{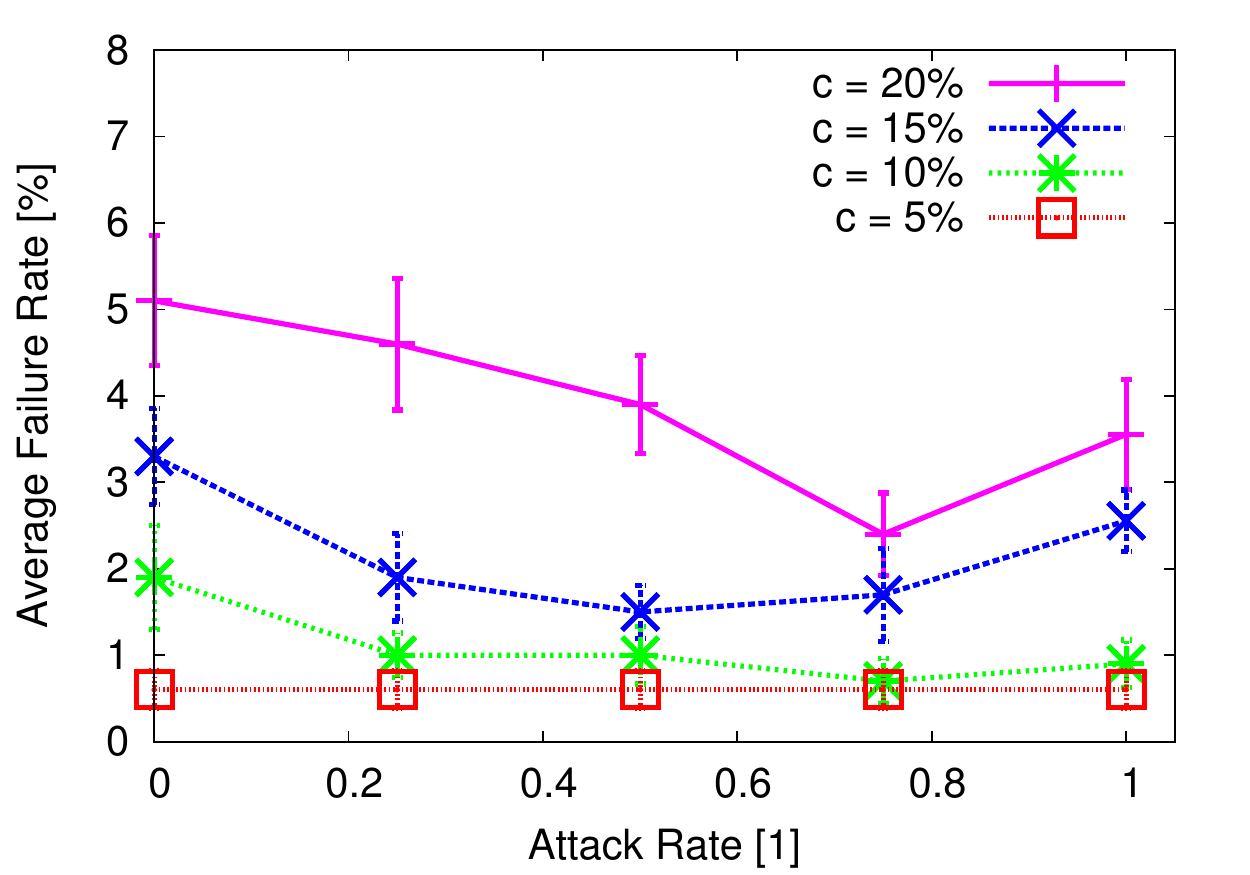}}
 \caption{{\bf Kad.} Overall attack effectiveness with different attack
   rates. Attackers perform best by attacking with low attack rates. }
      \label{fig:kad_churn_ar_graphs}
\end{figure}

\paragraphX{Kad.} We similarly examine overall attack effectiveness in
Kad. Figure~\ref{fig:kad_churn_ar_graphs} shows our simulation
results. Kad has very different results
(Figure~\ref{fig:kad_churn_areg}) from Halo due to routing table
pollution in the attacker model.
In particular, for Kad with colluding fraction $c=20\%$, the failure
rate is 80\% when the attack rate is $a=0.0$ and drops to 54\% when
$a=1.0$. The high failure rate with no manipulation of lookups ($a=0.0$)
is due to the effectiveness of routing table pollution against Kad. As
the attack rate increases, routing table pollution decreases, which
leads to the drop in failure rates.
This occurs because, when the attacker manipulates lookups, the results
returned by attacker nodes are always attacker nodes and are generally
further away from the target than those returned by honest nodes. So
whenever both attacker nodes and honest nodes are responding to a
lookup, the attacker nodes not only fail to manipulate the lookup result
but also do not appear in the later stages of the lookup process. This
reduces the malicious nodes' chances of being opportunistically added to
$k$-buckets.

In comparison, Figure~\ref{fig:kad_churn_a} shows how effective \ksys is
to counter the advanced route subversion attack described in
Section~\ref{sec:kad-reds}. The attackers gain a slight advantage with
lower attack rates. The reputation system, however, severely limits the
growth of the average failure rate by curbing routing table
pollution. \ksys maintains a failure rate of no more than 5.1\% for all
attack rates with $c=20\%$ or less, which is a 93\% improvement over
Kad.

\subsection{Comparison of subsearch failure reasons}

Next, we analyzed how and when subsearches (these are the redundant
lookups that together form the overall search) fail under the different
algorithms. A more indirect way of comparing different algorithms and
showing how the algorithms improve the failure rate is by looking at why
searches fail.

\begin{figure}[t!]
\centering \subfigure[{\bf Regular Halo.} As the colluding fraction
  increases, more subsearches fail because of a bad node in the lookup
  path.]{
    \label{fig:failure_regular}
    \includegraphics[width=\scalefac\columnwidth]{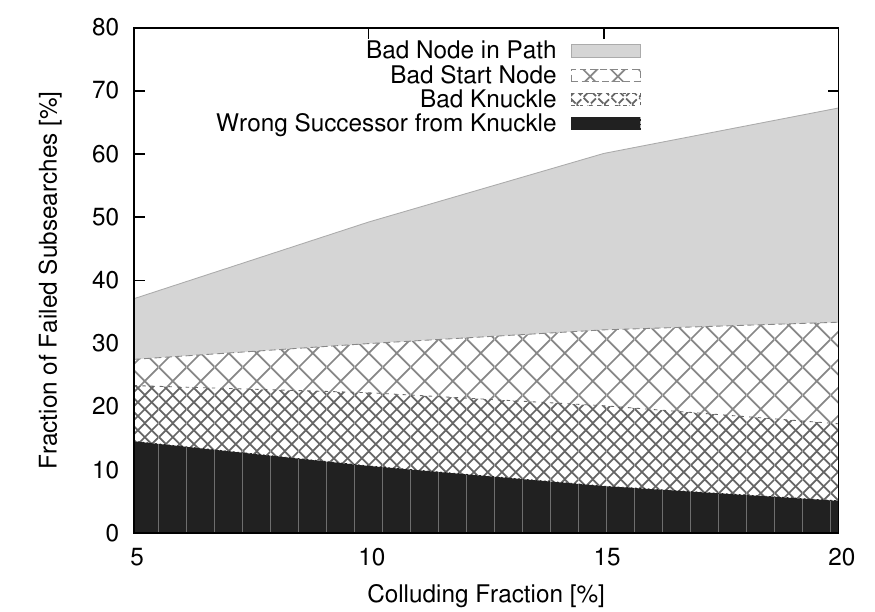}}
\hspace{0.5cm} \subfigure[{\bf A-Boost.} Compared to regular Halo,
  A-Boost has fewer failures due to a bad node in the path, showing that
  it is indeed routing around malicious nodes.]{
    \label{fig:failure_aboost}
    \includegraphics[width=\scalefac\columnwidth]{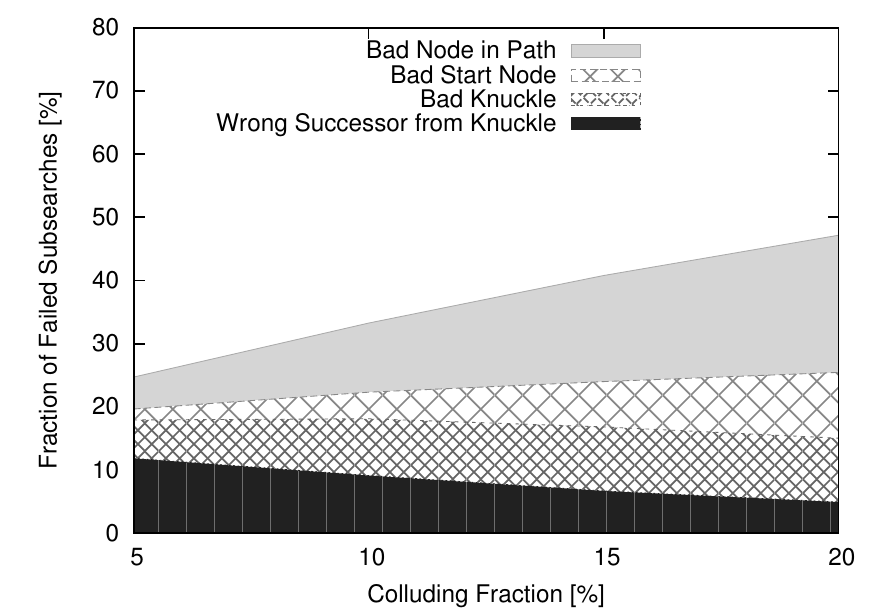}}
\subfigure[{\bf Collaborative boosting.} Even when the colluding
  fraction increases, failures due to a bad node in the lookup path
  increase much more slowly than in regular Halo or A-Boost.]{
    \label{fig:failure_collaborative}
    \includegraphics[width=\scalefac\columnwidth]{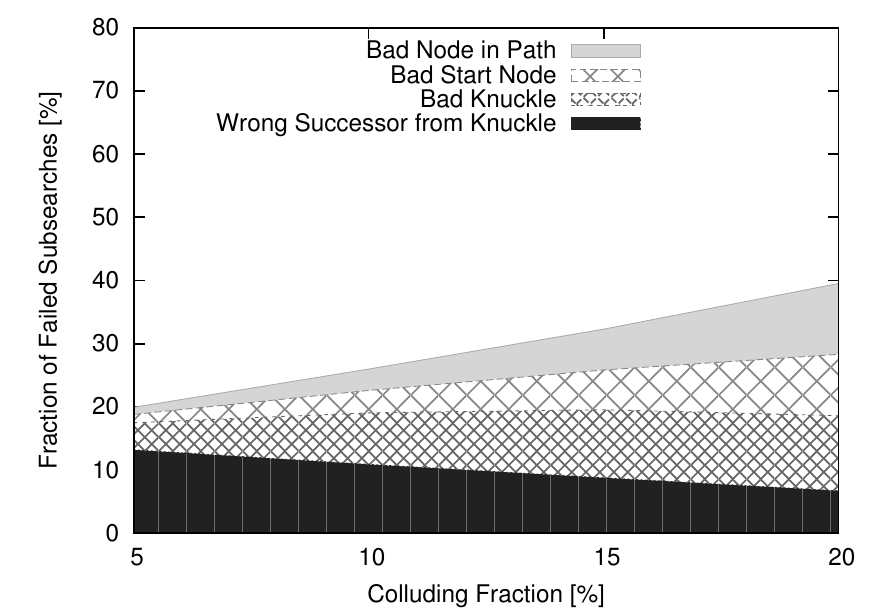}}
 \caption{{\bf Halo Failures.} What percentage of subsearches fail and
   for what reasons.}
      \label{fig:failure_graphs}
\end{figure}

Figure~\ref{fig:failure_graphs} shows the failure reasons for regular
Halo in Figure~\ref{fig:failure_regular} and collaborative boosting in
Figure~\ref{fig:failure_collaborative}. For both regular Halo and
collaborative boosting there are certain failures which cannot be
avoided.  For example, a knuckle may return the wrong successor node,
as this is an inherent limitation of Chord's structure, where a knuckle
for a particular exponential offset simply does not exist (25\% of the
time)~\cite{halo}. Other failures, however, can be reduced using
reputation information. The main failures for both regular Halo and
collaborative boosting are: a lookup hits a bad node in the lookup
path, the starting node for a knuckle-search was already a colluder, or
the knuckle that was found was a colluding node.  For regular Halo as
shown in Figure~\ref{fig:failure_regular}, the most important failure
reason is that a lookup hits a colluder node in the lookup path, which
subverts the search. As the colluding fraction increases, the number of
subsearches that fail because of a bad node in the path increases as
well.  For collaborative boosting as shown in
Figure~\ref{fig:failure_collaborative}, it is clearly visible that
collaborative boosting routes around bad nodes and thus has a much
smaller percentage of subsearches that fail due to a colluding node in
the lookup path. The percentage increases as the colluding fraction
increases, but it stays at a much smaller percentage compared to
regular Halo. Likewise, collaborative boosting is also more successful
at picking good nodes as starting nodes by exploiting reputation
information.  This shows that reputation information is indeed useful
in avoiding bad nodes in the lookup paths, by sending lookups through
parts of the Chord network which are known to be reliable.

\subsection{Effectiveness of shared reputation}\label{sec:results-shared}
In these experiments, we explore whether 
sharing reputation values can help lower the failure rate. We again
simulate malicious nodes attacking at a rate of 1.0, for different
fractions of colluding nodes, and see how the failure rate evolves
We also look at different ways of calculating shared reputation:
average, median, and the \sharedrep algorithm described in
Section~\ref{sec:reds:sReDS}.
The values returned by malicious nodes are furthermore calculated to
maximize the probability that the value is accepted by the requesting
node, by taking into account the shared reputation algorithm used. This
maximizes the attackers' advantage.

\begin{figure}[t!]
\centering 
    \includegraphics[width=\scalefac\columnwidth]{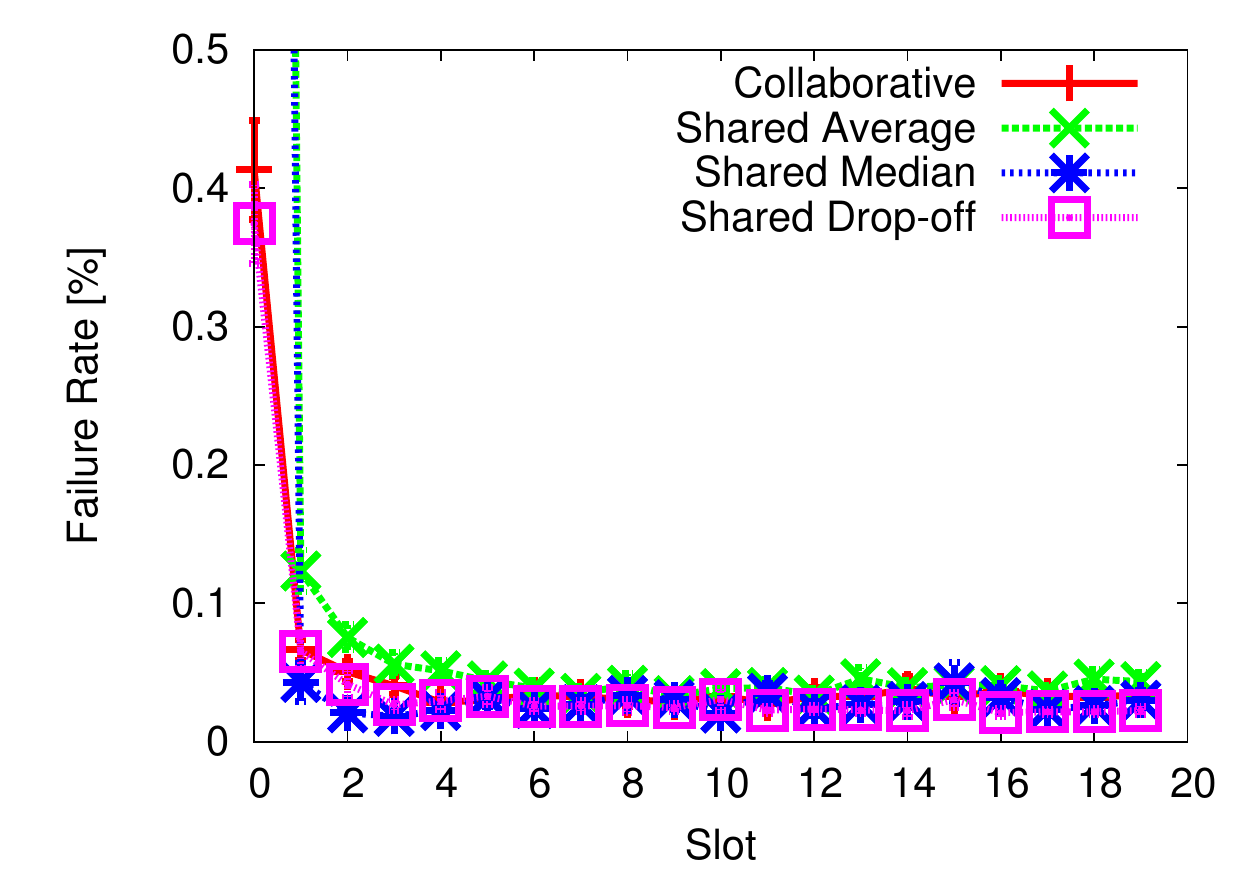}
 \caption{Shared reputation ($c=10\%$) at best performs about the same
   than normal collaborative mode (within the margin of error).}
      \label{fig:shared0.1}
\end{figure}

Figure~\ref{fig:shared0.1} shows the results for 10\% colluding
nodes. We see that while there is a slight difference in the
convergence speed at the beginning of the simulation (median and
\sharedrep shared reputation converging slightly faster), the difference
is not statistically significant. For a higher fraction of
colluding nodes of 20\% and for attack rates lower than $a=1.0$ (not
shown), shared reputation fails to improve the failure rate over
collaborative in any scenario.

\paragraphX{New Nodes.} Next we study whether new nodes joining the
system can benefit from shared reputation.
Intuitively, shared reputation should be particularly useful for new
nodes joining an existing Chord network. A new node does not have any
reputation information yet, so asking other nodes in the network for
shared reputation helps a node to make routing decisions until it has
collected enough observations by itself. In this experiment, we test the
failure rate of nodes newly added to the Chord network, where those
nodes rely solely on shared reputation and collaborative boosting.

\begin{figure}[t!]
\centering
  \includegraphics[width=\scalefac\columnwidth]{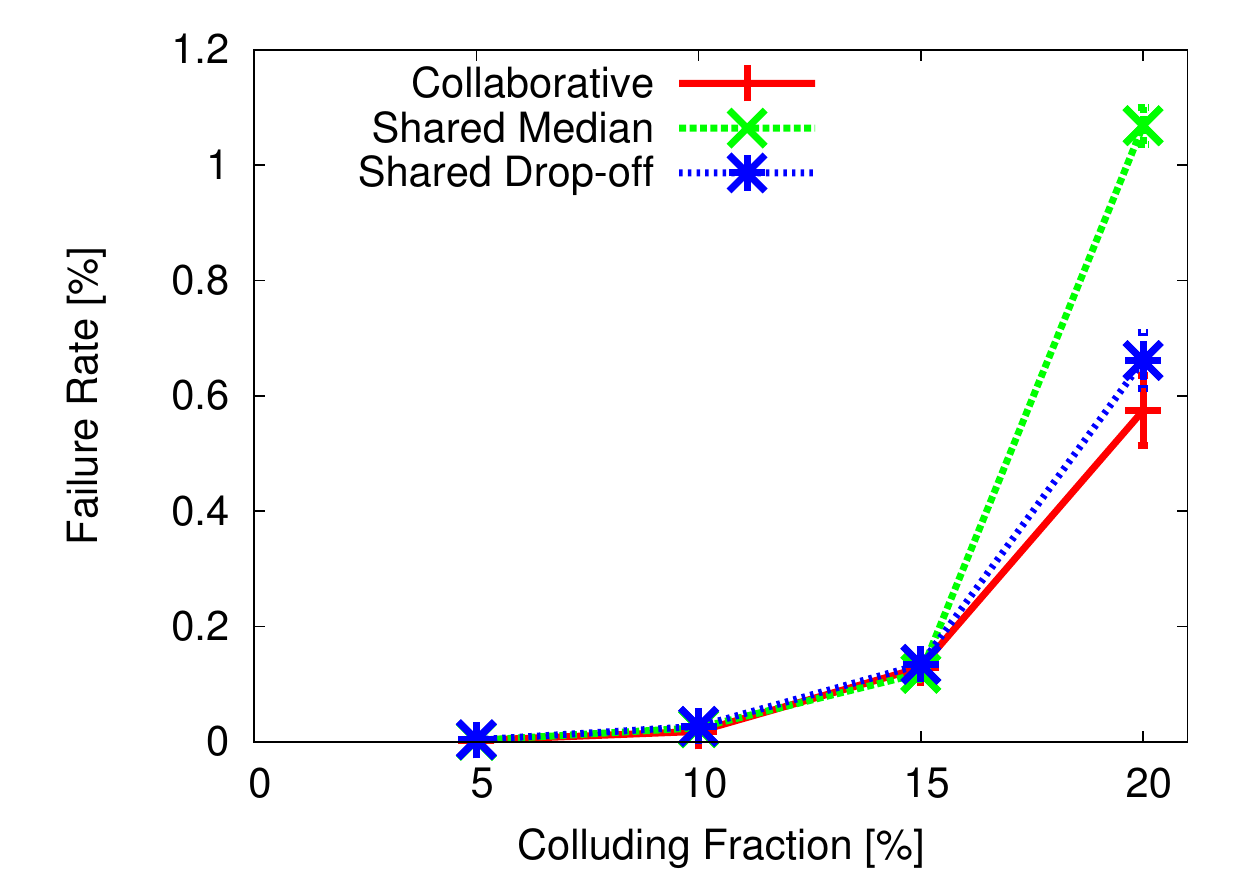}
  \caption{Using shared reputation increases the failure rate of newly
    joined nodes, because it is susceptible to reputation attacks by
    malicious nodes.}
  \label{fig:newnodes}
\end{figure}

The results do not bear out this intuition,
however. Figure~\ref{fig:newnodes} shows that using shared reputation
for newly added nodes does not improve the failure rate, and can even
worsen the failure rate for example for colluding fractions of
$20\%$. This can be explained by noting that using collaborative
boosting necessarily decreases (and cannot increase) the failure rate,
as each node only operates according to its first-hand observations. In
shared reputation, however, nodes become susceptible to slandering and
self-promotion of malicious nodes.

\section{Security Analysis}\label{sec:security}

In this section we present an analysis of the security of \sys against
various attackers. Since exploring these possibilities by fixing one or
a few parameters in simulation would be tedious and time consuming, we
analyze these situations theoretically. Based on our results, we
conclude that only oscillation attacks and targeted attacks on keys are
serious threats to \sys~--- we analyze oscillation attacks in this section
and show their effectiveness is limited. Targeted attacks are a further
challenge and we plan to address them in future work. We also examine a
novel attack against shared reputation in ReDS.

\subsection{Oscillation Attack} \label{sec:security:oscillation}

In an oscillation attack the attacker follows a strategic approach,
alternately acting as a benign node and then a malicious node. By behaving as an
honest node, the attacker increases its reputation scores in order to increase
the probability of being selected in future lookups, while performing
malicious activities in later periods. 
This attack can be especially
dangerous for \sys when lookups are made in recursive mode, since this adaptive behavior is hard to observe when
making indirect observations about the performance of lookup paths
beyond the first hop.


We now analyze the effectiveness of the oscillation attack and show that it
has  limited ability to undermine the system, especially over
time. The intuition of our finding is that the attacker must
either lose opportunities to attack while rebuilding his reputation
score or maintain a low reputation score and continually lose
opportunities to attack.

Although in the rest of the paper we study a simpler version of \sys, we
explore a more general version of \sys in this analysis. In particular,
we leverage an exponentially weighted moving average (EWMA) to track
nodes' scores with more emphasis on recent activity. The reputation
score ($s_{i+1}$) of a given node just before lookup $i+1$ is given by:
$s_{i+1} = \alpha r_i + (1-\alpha) s_i$,
where $r_i$ is the result of the lookup and $\alpha$ is the weight given
to the most recent results. We also allow the node to select a peer from
the $k$-bucket in a way that balances exploration (trying other nodes)
and exploitation (making use of the known scores). To do this we set
the probability of selecting the attacker node $a$ (let us call this
event $A$), who has score $s(a)$, as:
\[Pr[A] = \frac{s(a)^\beta}{\sum_{j \in k-bucket} s(j)^\beta},\]
where $\beta$ is a weighting parameter. These two generalizations allow
us to explore and understand the impact of these design choices in
analysis.


For the analysis we focus on a single attacker node and make the
following simplifications:

\begin{smitemize}
\item We examine the case when there is exactly one attacker and one
  honest node in a $k$-bucket.
\item The honest node's reputation score is fixed at $s_h$.
\item We do not consider churn.
\item When the attacker acts as a benign node, its reputation score is
  not affected by the malicious activities of any other nodes in the
  lookup path.
\end{smitemize}

\begin{figure}[tb]
  \centering
  \includegraphics[width=\scalefac\columnwidth]{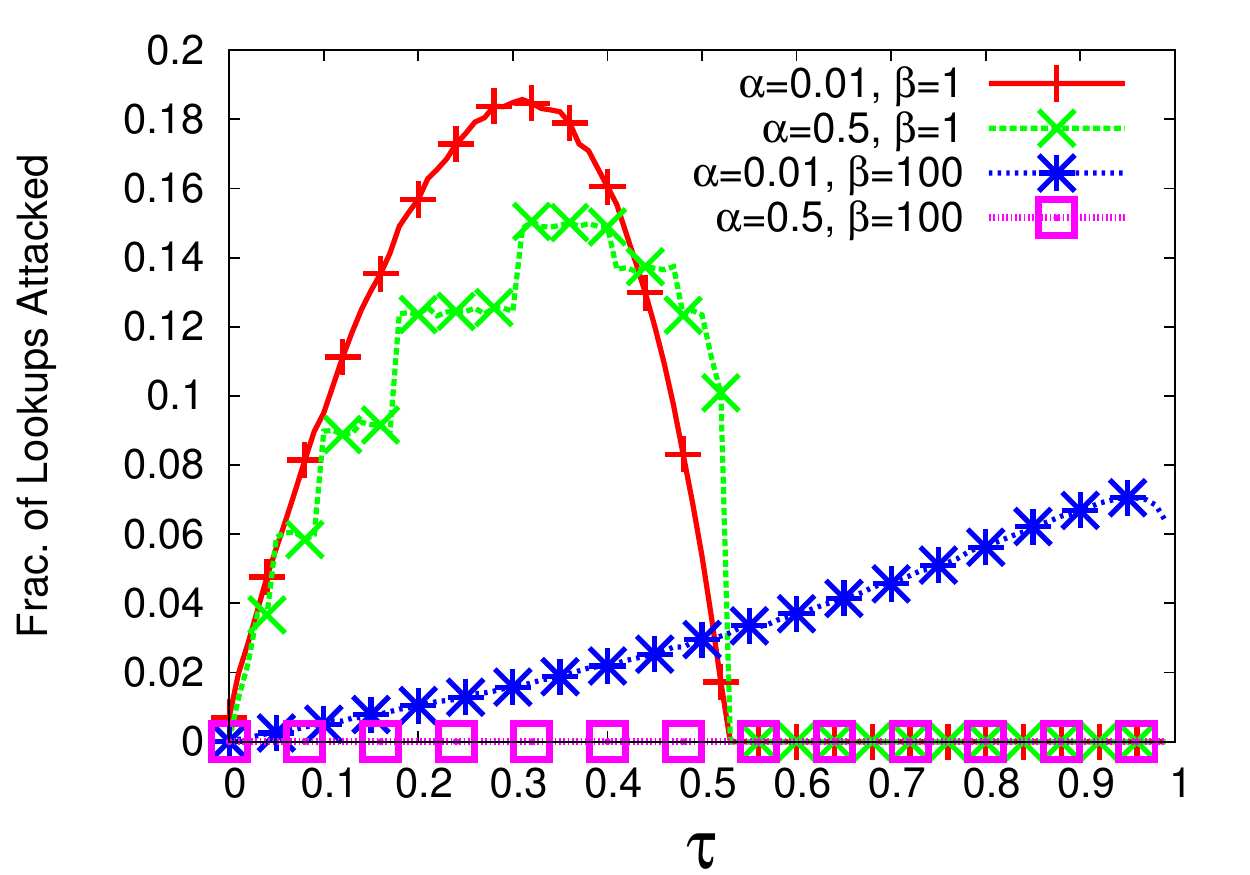}
  \caption{{\bf Analysis: One Threshold:} For varying $\tau$ and four
    combinations of $\alpha$ and $\beta$, the fraction of lookups
    attacked.}
  \label{fig:analysis:oscil-hard}
\end{figure}

%
The first two assumptions are for simplicity; our analysis generalizes
to various $k$-bucket populations and moderate fluctuations in the
honest node's score. By not considering churn we lose out on the
attacker's remaining opportunity to get lookups to attack. We evaluate
with churn in our extensive simulations in Section~\ref{sec:eval}. The
fourth assumption is the best strategy for our attacker. By coordinating
his malicious nodes to attack all at the same time, he only risks losing
reputation in an attack when he is also maximizing his chance to
modify a lookup result. Thus, the oscillation attack is a global
strategy.

To make the analysis tractable, we examine a limited set of possible
functions for the attacker to select the probability of attacking a
lookup: {\em one threshold}, {\em two thresholds}, and {\em
  probabilistic}. We examine each of these in turn.

\paragraphX{One Threshold.}
We first consider a threshold $\tau$ in which the probability of attack
on lookup $i$ is $p_i=1$ when $Pr[A]_i >= \tau$ and otherwise
$p_i=0$. This captures the intuition that the attacker should attack
when it is being selected often enough to have an impact and otherwise
it should rebuild his score.

The main metric we employ, and that our attacker seeks to maximize, is
the expected number of lookups that the attacker can attack
($E[attacks]$) given the total number of lookups $L$ that the user
performs through the $k$-bucket of interest. This can be written as:
\[E[attacks|L] = \sum_{i=1}^{L} Pr[A]_i \times p_i.\]
Since $p_i$ depends on $Pr[A]_i$, and each round's behavior depends on
the results of the prior rounds, we did not seek a closed-form
solution. Instead, we developed a simple numerical simulation of the
above formula for a range of values of $\tau$, $\alpha$, and $\beta$. We
examine the effect of $\tau$ for select values of $\alpha$ and $\beta$,
as shown in Figure~\ref{fig:analysis:oscil-hard}. We use $\alpha=0.01$
as a {\em slow-learning} model, emphasizing longer histories, and
$\alpha=0.5$ as a {\em fast-learning} model, emphasizing recent
behavior. Similarly, we use $\beta=1$ as a {\em lightly biased} model,
emphasizing exploration among $k$-bucket members, and $\beta=100$ as a
{\em heavily biased} model, emphasizing exploitation of
knowledge. Figure~\ref{fig:analysis:oscil-hard} shows that the attacker
can choose $\tau$ to attack a substantial fraction of lookups in both
lightly biased models and in the slow-learning, heavily biased model. In
these models the attacker can identify a peak at which $\tau$ is
optimal for the model. However, we also see that the fast-learning,
heavily biased model is very effective against this attacker, with
exactly one attack at all values of $\tau$. In this model,
$E[attacks]=1$, i.e. the attacker effectively never gets selected after
the first attack. This is  similar to the model that we use in \sys.

\begin{figure}[tb]
  \centering
  \includegraphics[width=\scalefac\columnwidth]{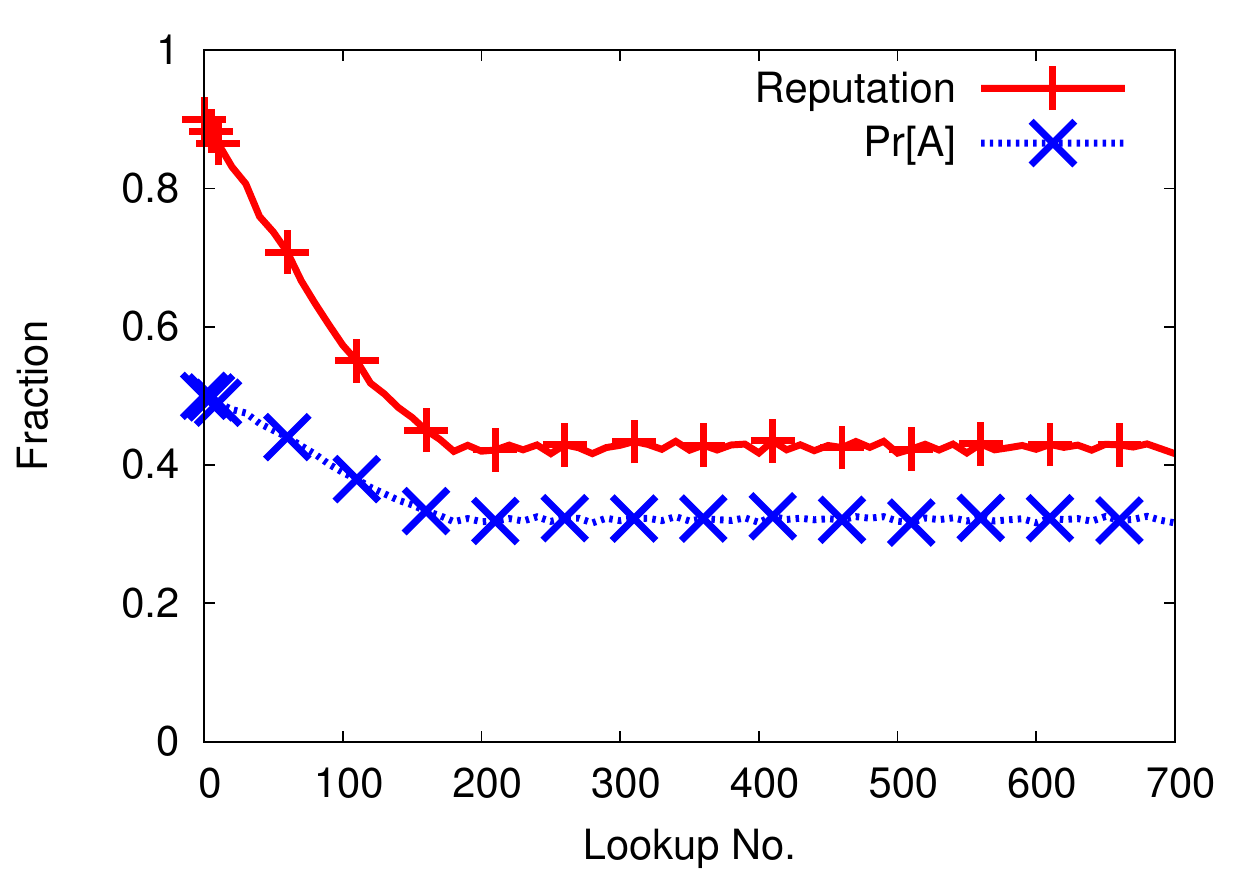}
  \caption{{\bf Analysis: One Threshold:} The reputation scores and
    $Pr[A]$ as the lookups proceed for $\alpha=0.01$ and $\beta=1$.}
  \label{fig:analysis:oscil-hard-best}
\end{figure}

To further break down  how the attacker modulates its behavior, we
examine the first few hundred lookups in the slow-learning, lightly
biased model in Figure~\ref{fig:analysis:oscil-hard-best}. We chose this
model with $\tau=0.32$ to show the best case for the attacker.
We see that the attacker's reputation score steadily
declines until oscillating around 0.42. The probability of being used
($Pr[A]$) similarly declines until oscillating around 0.32. The
oscillation attack is quite limited.

\paragraphX{Two Thresholds.}
Oscillating behavior may occur over longer time scales. To examine this,
we extend the threshold model to include a lower threshold $\tau_1$ and
an upper threshold $\tau_2$. The attacker will set $p=0$ whenever $Pr[A]
\leq \tau_1$, i.e. the attacker's reputation score has dropped too much
to be selected very often. He will set $p=1$ whenever $Pr[A] \geq
\tau_2$, i.e. the attacker has built up sufficient reputation to attack
again. The key question is how the attacker will set the thresholds
$\tau_1$ and $\tau_2$.

\begin{figure}[tb]
  \centering
  \includegraphics[width=\scalefac\columnwidth]{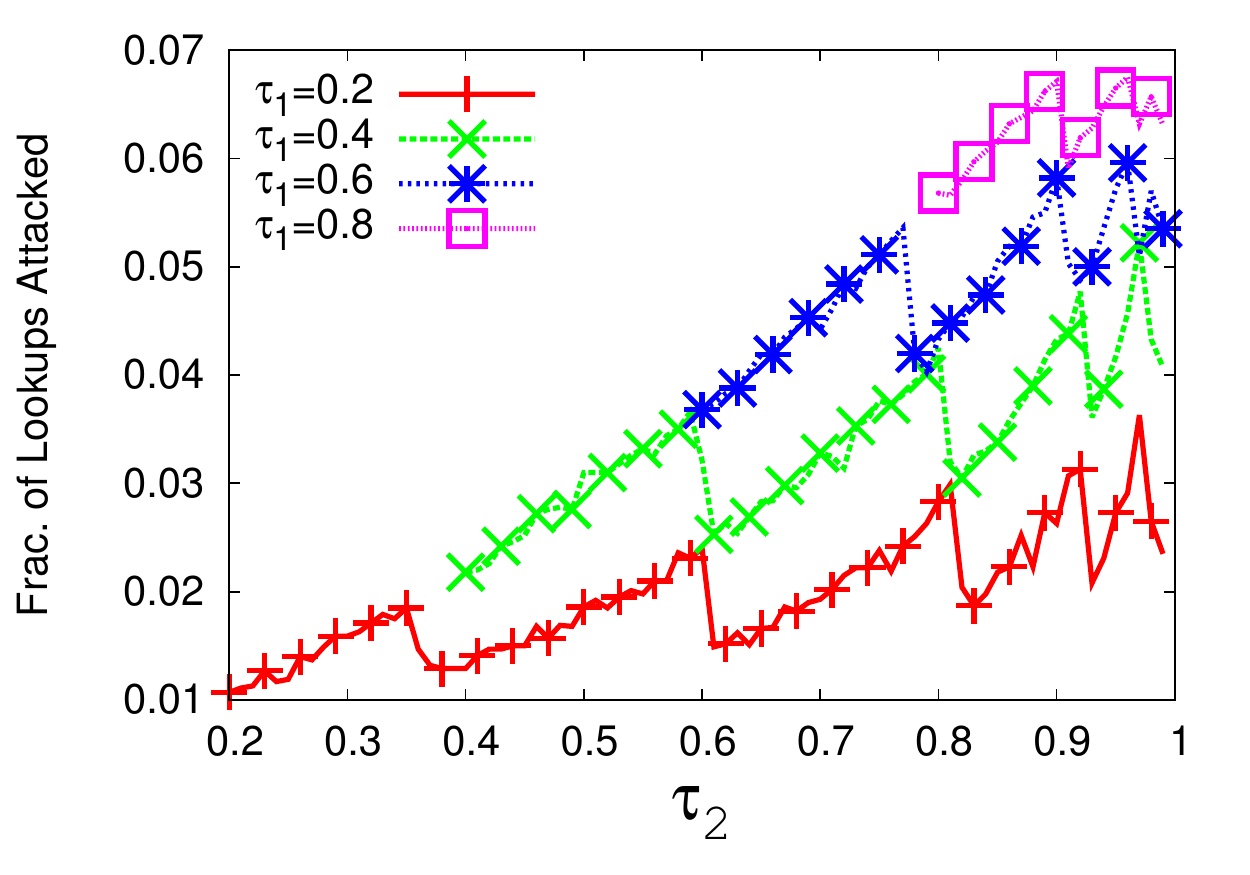}
  \caption{{\bf Analysis: Two Thresholds:} For varying $\tau_1$ and
    $\tau_2$, the fraction of lookups attacked.}
  \label{fig:analysis:oscil-range}
\end{figure}

In Figure~\ref{fig:analysis:oscil-range} we see the attack rates for
lookups in the slow-learning, heavily biased model. We have similar
results for each model as with the one threshold attacker. First we
note that, in this model, the attacker is never able to attack more than
7.1\% of lookups. Further, the attacker's best strategy is to keep his
range of scores quite high, requiring him to behave honestly for most
lookups.
In the fast-learning, heavily biased model, the attacker can never
attack more than one lookup.

\begin{figure}[tb]
  \centering
  \includegraphics[width=\scalefac\columnwidth]{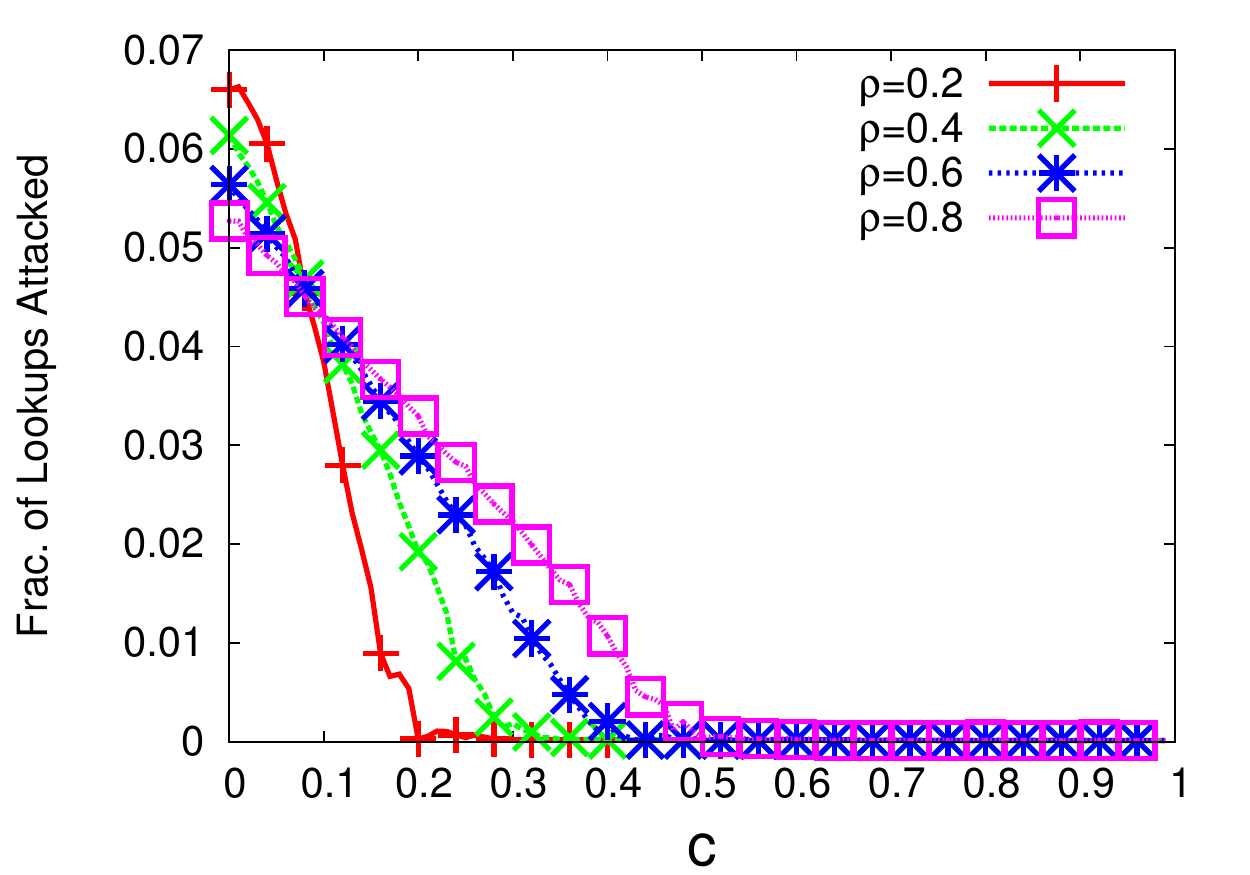}
  \caption{{\bf Analysis: Probabilistic:} For varying $\rho$ and $c$,
    the fraction of lookups attacked.}
  \label{fig:analysis:oscil-prob}
\end{figure}

\paragraphX{Probabilistic.}
Since the attacker can also employ a probabilistic attacking strategy,
we also examine a probabilistic version of the oscillation attack. We
let the attacker's probability $p$ of attack for a given lookup $i$ be:
\[p_i = \rho (Pr[A]_i - 0.5) + c\]
for attacker-chosen constants $\rho$ and $c$. Although this function is
linear, it covers a wide range of possible attacker policies.
Figure~\ref{fig:analysis:oscil-prob} shows the change in number of
lookups attacked in the slow-learning, heavily biased model for varying
$\rho$ and $c$. As with the other attacker models, the attacker has very
limited success (again, he can only attack at most 7.1\% of
lookups). Additionally, the fast-learning, heavily biased model still
only allows for one attack.

In sum, in all three attacker models, the oscillation attack provides
little to no advantage to the attacker.

\subsection{Other attacks on first-hand observations}

In \sys a node maintains its own reputation tree for each node in its
$k$-buckets. Other than an oscillation attack, there are several ways an
attacker might try to manipulate first-hand observations. White-washing,
bootstrapping, and targeted attacks are three such attacks that we
briefly discuss in this section.

\paragraphX{White-washing Attacks.} In a white-washing attack, a node
leaves and rejoins the system to get a better reputation score. This
attack can be partially mitigated by having nodes cache
reputation values for nodes that have left, up to a memory limit, 
and by setting a low initial reputation score for new nodes that
discourages this attack. 
Through a white-washing attack the attacker
could also attempt to gain a higher reputation score in the joining
round than it would have by staying in the system and behaving honestly
in a standard oscillation attack. From the above analysis of the
oscillation attack, we could identify the expected score at a time when
the attacker's reputation reaches its nadir (say, $s_{low}$) and the
benefit of white washing would be greatest. We should then set the
initial reputation of joining nodes to $s_{low}$ to remove the incentive
for white washing as long as the attacker behaves optimally in the
oscillation attack.

\paragraphX{Bootstrapping Attacks.} In the beginning phase of the
system, we do not have enough observations for nodes to build their own
reputation scores. In this phase we give each node an initial
reputation score, and the probability of a node being selected for the
first lookup from a given $k$-bucket is $1/k$. If the node returns a bad
result, then the requesting node immediately switches to another node in the
$k$-bucket, limiting the effect of an all-out attack in the early phases
of the system. With time, we get the required observations and peers can
distinguish between the honest and malicious nodes in their
$k$-buckets.

\paragraphX{Targeted Attacks on Keys.} Attackers in \sys may
also try blocking access to a specific resource, or provide a malicious
version of the resource, without attacking other lookups in the system,
i.e., the attacker only manipulates lookups for a specific target key
$t$. This is more challenging for \sys than generic attacks because it
can only be observed when the desired resource is being requested. We
believe that limited tracking of attacked keys may be possible, but we
leave further exploration to future work.

\paragraphX{Targeted Attacks on Users.} Similarly, an attacker may be
interested in preventing a specific peer from accessing resources in the
system. Since \sys is most effective with the collaborative help of
other nodes, the benefits of \sys are limited against this attack.


\subsection{A use-based attack on shared reputation}\label{sec:shared-use}

We now consider attacks on shared reputation. Despite the relative
resilience of the \sharedrep scheme, it is vulnerable to a novel attack
that greatly affects the possibility of shared reputation in ReDS. This
{\em use-based attack} works against any ReDS system in which a given
finger is used more by some knuckles than others. The attacker seeks to
limit the loss of reputation from attacks while attacking as many
lookups as possible. The attacker can achieve this by
attacking the lookups from its knuckles who use his node as a finger
more while not attacking lookups from other nodes. When the joint
knuckles share reputation information about this malicious finger, they
will have conflicting scores. The attacker's goal is to arrange its
attacks so that the low scores are mostly ignored by other nodes.


We now describe a use-based attack in detail as applied to Halo-ReDS
with shared reputation. A version of this attack should also work
against Kad-ReDS with shared reputation, due to the XOR metric, or against any ReDS system in which a
large fraction of lookups go through just a few fingers. For simplicity,
we assume that each node in the Halo DHT performs the same number of
lookups. The assumption is valid when peers perform a large number of
lookups, and the probability of a given peer to initiate a lookup follows
uniform distribution. With non-uniform distributions of lookup rates,
the attack should have the same results on average.

In the use-based attack the attacker node acts as a malicious finger
for $m$ of its $k$ knuckles and as an honest node for the remaining
$k-m$ knuckles. An attacker node with ID $a$ attacks the $m$ most
distant knuckles, as these knuckles use node $a$ to cover a larger
fraction of the ID space. In particular, $a$ performs maliciously for
the knuckles having ID $a-2^{log(n)-i}$, where $i=1, 2,\dots ,m$. Given
$l$ lookups using node $a$, we estimate that the number attacked on
average will be
$\sum_{i=1}^{m} \frac{l}{2^i} = l \left(1-\left(\frac{1}{2}\right)^m\right)$.

\begin{figure}[t]
  \centering
  \includegraphics[width=0.7\columnwidth]{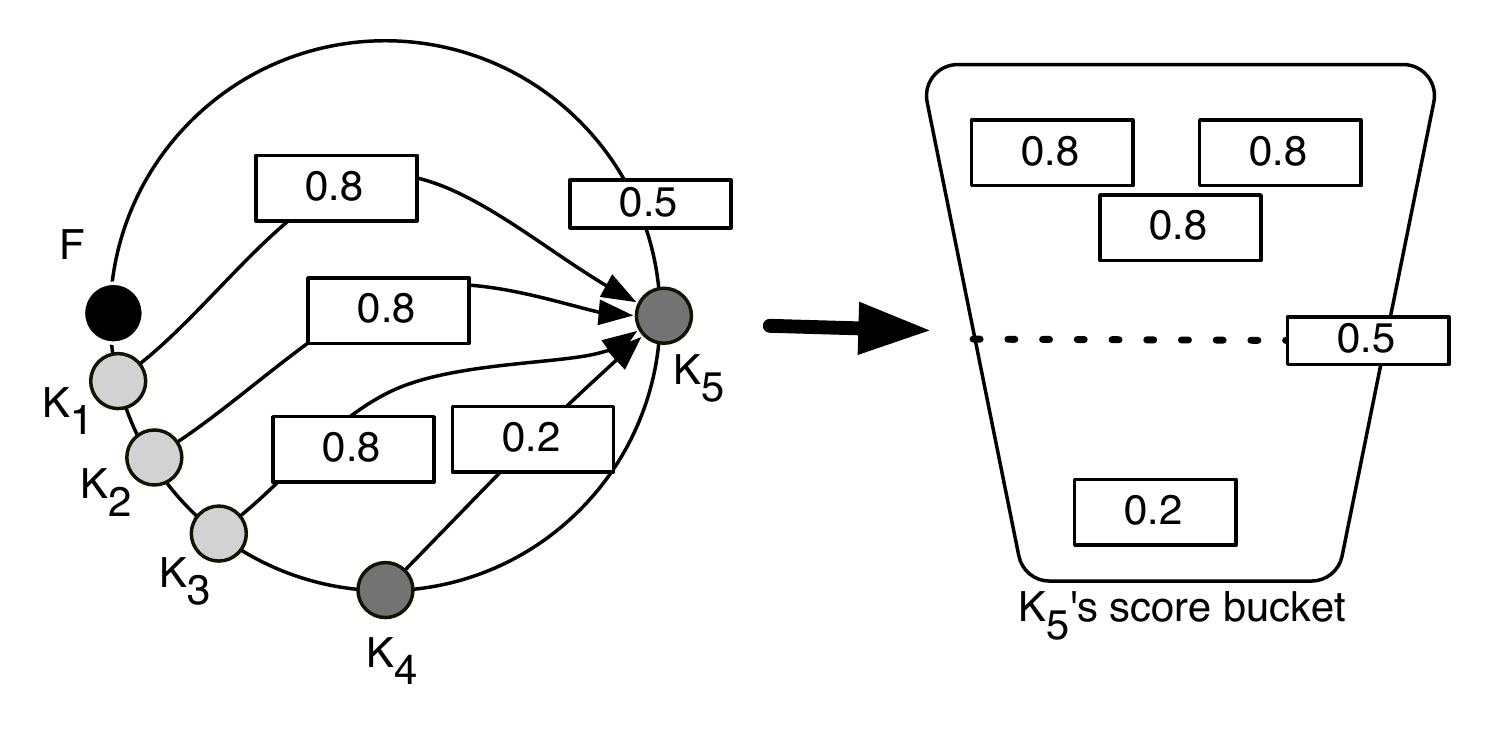}
  \caption{{\bf A use-based attack.} $F$, a malicious finger, attacks
    lookups from knuckles $K_4$ and $K_5$ but not those from $K_1$,
    $K_2$, and $K3$. Scores are reported to $K_5$, whose score bucket is
    shown on the right.}
  \label{fig:use-based-diagram}
\end{figure}

We show an example of the attack in
Figure~\ref{fig:use-based-diagram}. $F$ is a malicious finger with
knuckles $K_1$ to $K_5$. Here $m=2$, meaning that lookups from $K_4$ and
$K_5$ are being attacked, accounting for 75\% of the lookups through
$F$. $K_5$ is a new node with reputation score 0.5 for $F$, whereas
$K_4$'s score of 0.2 for $F$ reflects $F$'s attacks on its lookups.
We show the reputation scores sent to $K_5$, which include three scores
of 0.8 from knuckles $K_1$ to $K_3$ and 0.2 from $K_4$.  Using the
median, the score will be 0.8, while using \sharedrep, the expected
score is 0.75. In either case, the finger can thus attack many lookups
while retaining a high reputation.

We further study the use-based attack in a simple simulator of the
\sharedrep scheme, using $10000$ nodes and $10000$ lookup
operations. Since node $a$ may not always behave the same to a given
knuckle, we define two parameters. For the $m$ knuckles for which $a$
acts maliciously, let $f$ represent the percentage of lookups through
$a$ that fail. Let $s$ as the percentage of successful lookups through
$a$ for the $k-m$ of knuckles for which $a$ acts honestly.

For example, if $s=80$\% and $f=80$\%, the victim knuckles give $a$ a
score of $0.2$ and the other $k-m$ knuckles, $0.8$. 
In Figure~\ref{fig:sreds1} we consider the shared reputation
scoring of the $m$ knuckles when using all $k$ scores. When $m=1$,
$s=80$\%, and $f=80$\% the lone victim knuckle 
uses $0.8$ as the shared reputation score of $a$ in $p=98.6$\% of
cases. At the same time, he can attack 50\% of all lookups going through
it.
If an attacker acts maliciously for more knuckles, it causes more
lookups to fail, but its credibility is decreased to those
knuckles. Thus, the value of $p$ decreases as we increase
$m$. Figure~\ref{fig:sreds1} shows that for $m=5$, we get $p=43.1$\%,
while the attacker can attack 78\% of lookups.

\begingroup
\setlength\belowdisplayskip{0pt}
\begin{figure}[!t]
\centering
\includegraphics[width=\scalefac\columnwidth]{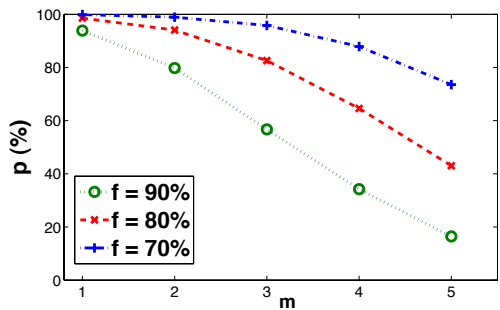}
\caption{Percentage of cases ($p$\%) in which a given attacked knuckle
  computes the reputation score of the attacker finger as 0.8
  ($s=80$\%)}
\label{fig:sreds1}
\end{figure}
\endgroup

In sum, the use-based attack enables the attacker to attack a majority
or large fraction of lookups while still getting a good reputation score
most or nearly all of the time.


\paragraphX{Countermeasures.} We first note that the \sharedrep scoring
scheme may not be the best suited to stop the attack, as it is designed
mainly to resist slandering and self-promotion attacks. Basic schemes,
however, fare even worse. Using the median, the attacker would be able
to attack half of its knuckles and still attain an excellent reputation
score. For 10,000 nodes, this means the attacker could attack six out of
13 knuckles, covering 98.4\% of lookups and have a perfect reputation
score. Average is better, but is much more vulnerable to slandering and
self-promotion. One could note that in \sharedrep, the node is ignoring
its own score to its detriment. Making the score more centered on the
node's own local score, however, means not obtaining any significant
benefit from sharing reputation over only using first-hand observations.

One could attempt to design a scheme specifically to counter this
attack, but it must also resist slandering and self-promotion
attacks. For example, one could weight the scores of distant knuckles
more heavily than nearby knuckles to reflect greater use by distant
knuckles. Unfortunately, weighting the scores of any knuckles more
heavily gives them greater power to perform slandering or
self-promotion. 
Another countermeasure is to use a DHT in which all fingers are used
equally. This suggests that Salsa, in which all local contacts are used
equally~\cite{salsa}, is more suitable for shared reputation.
Considering the combined effect of the
use-based attack, the limited benefits shown in
Section~\ref{sec:results-shared}, and the overhead of shared reputation,
we recommend against shared reputation in ReDS.





\section{Related Work}\label{sec:related}

In a paper about secure routing in peer-to-peer networks (focusing on
Pastry, but generalizable to other protocols), Castro et
al.~\cite{castro02secp2p} argue that secure routing requires secure
assignment of identifiers, secure routing table maintenance, and secure
message forwarding. Secure assignment of identifiers is done through the
use of a certificate authority (CA) which binds identifiers to IP
addresses. Solving the problem of secure routing table maintenance
requires modification of the Pastry protocol to introduce additional
constrained routing tables. Lastly, secure message forwarding is
approached by detecting failed routes and then applying route
diversity. Route diversity is achieved by forwarding multiple messages
until they reach a node which has the target node for a key in its
neighbor set. We argue that \sys can be used effectively for
any system designed along the lines of Castro et al.'s secure routing
primitive.

In a scheme focusing on Chord, Harvesf et
al.~\cite{DBLP:journals/tdsc/HarvesfB11} describe an algorithm using
replica placement to improve routing robustness in a peer-to-peer
network. Specifically, by placing several replicas of a key uniformly
around the Chord network, disjoint routes to the individual replicas are
created, which makes it likely that at least one search for one of the
replicas will use a route with no compromised nodes. This approach of replication is orthogonal to our work (indeed Kad too uses multiple `replica roots'). ReDS ensures that searches for each replica will succeed with higher probability, and thus fewer replicas need to be retrieved, or fewer replicas are needed in the first place.


Mickens et al. propose a system called
``Concilium''~\cite{DBLP:conf/dsn/MickensN07}, which attempts to distinguish
between malicious behavior and network problems and assigns blame to
nodes if they are found to subvert searches. It also depends on secure
identifiers (e.g., using a CA) like the scheme by Castro et
al.~\cite{castro02secp2p}.
Concilium focuses more on diagnosis and identifying malicious nodes. It
requires nodes to perform network tomography as well as propagate
`Blame' messages downstream to identify malicious nodes, both of which
require coordination. \sys does not try to implicate and remove bad
nodes, but simply avoids them, thereby limiting the cost of false
positives and allowing for fast decisions. \sys also does not require
nodes to coordinate reputation information among themselves, reducing
overhead and complexity.

Malicious attacks in DHTs can be partially
addressed by using the concept of quorums. A quorum is a group of nodes
that effectively acts as an atomic unit, replacing individual peers in
the DHT.
Quorums consist of $O(\log n)$ nodes where $n$ is the total number of
nodes in the system.
There are several different approaches to create and maintain
quorums~\cite{quorumRef1, quorumRef2, quorumRef3, quorumRef4,
quorumRef5}.
Young et al. propose a quorum-based system~\cite{quorumRef5} that can
tolerate a large fraction of malicious peers~--- strictly less than
$1/3$-fraction of a quorum. We note, however, that if 10-20\% of the
nodes are attackers, a substantial fraction of quorums will be
controlled by attackers. \sys can thus improve outcomes for quorum-based
systems by applying reputation at the quorum level instead of the node
level.





\section{Conclusions}
\label{sec:conclusions}

We presented \sys, a reputation-based mechanism from improving the resilience of searches in deterministic and non-deterministic DHTs such as Halo (based on Chord) and Kad (based on Kademlia) against malicious nodes. We showed how information from failed searches can be used collaboratively to  avoid effectively malicious activity in the network. Our results improve significantly over  Halo and Kad,  showing that even exclusively local observations for reputation information can deliver large gains to the success rate when used collaboratively. We analyzed the potential for shared reputation mechanisms and a novel attack against shared reputation. We hope our work stimulates more research in reputation systems for structured peer-to-peer networks where structural information can be exploited for enhanced resilience against attackers.

\section*{Acknowledgments}

This material is based upon work supported by the National Science Foundation under Grant Nos. CNS-1117866, CNS-1115693 and CAREER award number CNS-0954133. We also thank John McCurley for his editorial help.


\bibliographystyle{abbrv}
\bibliography{main}


\end{document}